\documentclass[
 twocolumn,
 floatfix,
 a4paper,
 showpacs,
 showkeys,
 nofootinbib,
 reprint
 ]{revtex4-1}

\usepackage{epsfig}
\usepackage{latexsym}
\usepackage{xspace}
\usepackage[colorlinks=true,linktocpage=true,linkcolor=blue,citecolor=blue,allcolors=blue]{hyperref}
\usepackage[utf8]{inputenc}
\usepackage{indentfirst}
\usepackage{enumerate}
\usepackage{color}
\usepackage{todonotes}

\usepackage[caption=false,position=top]{subfig}

\usepackage{amsmath}
\usepackage{amssymb}
\usepackage[english]{babel}
\usepackage{url}

\newcommand*{\ham}{\mathcal{H}}

\newcommand{\bvar}{\mathbf}

\newcommand{\eq}[1]{\begin{align} #1 \end{align}}
\newcommand{\mean}[1]{\langle #1 \rangle}

\newcommand{\sNN}{\sqrt{s_{\rm NN}}}
\newcommand{\fsampler}{\texttt{FIST sampler}}

\begin{document}

%TC:ignore

\title{
Cooper-Frye sampling with short-range repulsion
}

\author{Volodymyr Vovchenko}
\thanks{\emph{Present address:} Physics Department, University of Houston, Box 351550, Houston, TX 77204, USA}
\affiliation{Institute for Nuclear Theory, University of Washington, Box 351550, Seattle, WA 98195, USA}
\affiliation{Frankfurt Institute for Advanced Studies, Giersch Science Center,
D-60438 Frankfurt am Main, Germany}
\affiliation{Nuclear Science Division, Lawrence Berkeley National Laboratory, 1 Cyclotron Road,  Berkeley, CA 94720, USA}

\begin{abstract}
This work incorporates the effect of short-range repulsion between particles into the Cooper-Frye hadron sampling procedure.
This is achieved by means of a rejection sampling step, which prohibits any pair of particles from overlapping in the coordinate space, effectively modeling the effect of hard-core repulsion.
The new procedure -- called the \texttt{FIST sampler} -- is based on the package \texttt{Thermal-FIST}. 
It is used here to study the effect of excluded volume on cumulants of the (net-)proton number distribution in central collisions of heavy ions in a broad collision energy range in conjunction with exact global conservation of baryon number, electric charge, and strangeness. The results are compared with earlier calculations based on analytical approximations, quantifying the accuracy of the latter at different collision energies.
An additional advantage of the new method over the analytic approaches is that it offers the flexibility provided by event generators, making it straightforwardly extendable to other observables.
\end{abstract}

\maketitle

%TC:endignore 

\section{Introduction}
\label{sec:intro}

The description of bulk observables in relativistic heavy-ion collisions is usually performed in the framework of relativistic hydrodynamics~(see e.g.~\cite{Bzdak:2019pkr,Shen:2020mgh} for an overview). This modeling incorporates the Cooper-Frye particlization stage, where the expanding QCD fluid is transformed into a gas of hadrons and resonances.
In most cases, hadron momenta are sampled from local Maxwell-Boltzmann distributions, with the possible inclusion of shear and bulk viscous corrections~\cite{Kisiel:2005hn,Pratt:2010jt,Shen:2014vra,Karpenko:2015xea,Bernhard:2018hnz}. 
Multiplicity distributions of the sampled hadrons usually correspond to Poisson statistics, implying independent particle emission.
More involved descriptions additionally implement a (micro-)canonical treatment of conservation laws, such as that of energy-momentum and QCD conserved charges~\cite{Becattini:2003ft,Becattini:2004rq,Schwarz:2017bdg,Oliinychenko:2019zfk,Oliinychenko:2020cmr}.
Proper treatment of conservation laws is relevant for observables involving event-by-event fluctuations~\cite{Begun:2004gs,Begun:2006jf,Braun-Munzinger:2020jbk,Vovchenko:2020kwg,Vovchenko:2021kxx}.

Physically, the system created at particlization corresponds to an ideal hadron resonance gas, possibly with non-equilibrium corrections due to shear and bulk viscosities and (micro-)canonical effects.
On the other hand, extensions of the Cooper-Frye procedure are necessary to incorporate any additional physics.
One commonly discussed extension of the ideal HRG is short-range repulsive interactions utilizing excluded volume~\cite{Yen:1997rv,Satarov:2016peb,Vovchenko:2020lju}.
For example, by introducing the excluded volume effect into baryon-baryon~(and, by symmetry, antibaryon-antibaryon) interaction~\cite{Vovchenko:2016rkn}, one can effectively model the presence of hard core in nucleon-nucleon scattering
and improve the description of several lattice QCD susceptibilities~\cite{Vovchenko:2017xad,Huovinen:2017ogf,Karthein:2021cmb,Bollweg:2021vqf}.
Another extension would concern the search for the QCD critical point, which would lead to large non-Gaussian fluctuations of the proton number~\cite{Stephanov:2008qz} requiring a generalized Cooper-Frye routine~\cite{Ling:2015yau,Pradeep:2022mkf}.

Previously, a method called subensemble sampler was introduced in~\cite{Vovchenko:2020kwg} to perform particlization of an \emph{interacting} hadron resonance gas and was used to study the influence of baryonic excluded volume on proton and baryon number cumulants in Pb-Pb collisions at LHC energies.
Although this method is rather generic, it requires partitioning the Cooper-Frye hypersurfaces into patches which should, on the one hand, be large enough to capture all the relevant correlations but, on the other hand, also be small compared to inhomogeneity scales across the hypersurface.
How to perform the partition into patches can be ambiguous.
The method has been used to study rapidity acceptance dependent observables at LHC energies, where the approximate longitudinal boost invariance is realized~\cite{Vovchenko:2020kwg}. Its application at lower collision energies, however, is more complex.
Instead, in Ref.~\cite{Vovchenko:2021kxx}, the proton cumulants at RHIC-BES were analyzed using an analytic approach incorporating baryon repulsion and conservation. 
However, analytic approaches lack the flexibility of event generators and are typically restricted to specific observables, such as (net-)proton number cumulants in a particular acceptance~\cite{Vovchenko:2021kxx}.
Therefore, it is advisable to incorporate the effect of short-range repulsion directly into the sampling procedure.

This work introduces the effect of short-range repulsion between particles into the Cooper-Frye particlization through a rejection sampling step, which prohibits any pair of particles from overlapping in the coordinate space.
This induces negative correlations between particles that are shown to be consistent with analytical expectations based on the excluded volume model.
Then, the method is used to study the effect of baryon repulsion on the cumulants of (net-)proton number distribution in central heavy-ion collisions in a broad energy range $\sNN = 2.4-2760$~GeV while simultaneously incorporating exact global conservation of baryon number, electric charge, and strangeness.

The paper is organized as follows.
The modeling of the short-range repulsion effect using the rejection sampling step is described in Sec.~\ref{sec:method}.
Section~\ref{sec:CooperFrye} describes how the effect is introduced into a multi-component hadron system at Cooper-Frye particlization. The method's application to heavy-ion collisions is presented in Sec.~\ref{sec:HIC}. Conclusions and outlook in Sec.~\ref{sec:summary} close the article.

\section{Method}
\label{sec:method}

\subsection{Probability of configurations}

Consider a uniform system of $N$ classical particles in volume $V$ that is in contact with the heat bath characterized by temperature $T$.
Interactions between particles are mediated by a (non-relativistic) pair potential $V(\bvar r_i,\bvar r_j)$.
The properties of the system are characterized by the canonical ensemble $(T,V,N)$.
The microscopic configuration of the system can be given by the set $\{\bvar r_i, \bvar p_i\}$ of coordinates and momenta of all particles in the system.
The probability of a particular configuration is determined by the Boltzmann factor, i.e.
\eq{\label{eq:prob}
P(\{\bvar r_i, \bvar p_i\}) \propto e^{-\frac{\ham\left(\{\bvar r_i, \bvar p_i\}\right)}{T}},
}
where $H\left(\{\bvar r_i, \bvar p_i\}\right)$ is the system Hamiltonian comprising the kinetic and potential energy terms:
\eq{\label{eq:H}
H\left(\{\bvar r_i, \bvar p_i\}\right) = \sum_{i=1}^N \varepsilon(\bvar p_i)  + \frac{1}{2} \sum_{i,j=1}^N V(\bvar r_i, \bvar r_j).
}
Here $\varepsilon(\bvar p_i)$ is the energy-momentum relation for the particle $i$, for example, $\varepsilon(\bvar p_i) = \sqrt{\bvar p_i^2 + m^2}$ for relativistic particles and $\varepsilon(\bvar p_i) = \bvar p_i^2 / (2m)$ for non-relativistic particles.

Based on the structure of the Hamiltonian in Eq.~\eqref{eq:H} it is clear that the probability $P(\{\bvar r_i, \bvar p_i\})$ factorizes into momentum- and coordinate-dependent parts.
The momentum distribution is given by the (non-)relativistic Maxwell-Boltzmann distribution. Thus, the sampling of particle momenta proceeds in the standard way.
However, the probability distribution of the coordinates of the particles is affected by the interaction potential $V(\bvar r_i,\bvar r_j)$. The unnormalized distribution density reads
\eq{\label{eq:Pcoord}
\tilde P ^{\rm coord}(\{\bvar r_i\}) = \prod_{i,j=1}^N e^{-\frac{V(\bvar r_i,\bvar r_j)}{T}}.
}

Without loss of generality, consider now that the system is placed in a cubic volume.
A sampling of momenta and coordinates can be performed via rejection sampling.
First, the coordinates are sampled uniformly from the cubic volume.
Then, the particle momenta are sampled from the Maxwell-Boltzmann distribution.
The sampled configuration is accepted with relative weight
proportional to $\tilde P^{\rm coord}(\{\bvar r_i\})$ in Eq.~\eqref{eq:Pcoord}.
For purely repulsive potentials, $V(\bvar r_i,\bvar r_j) \geq 0$, the maximum weight computed through Eq.~\eqref{eq:Pcoord} does not exceed unity, and thus the application of rejection sampling is straightforward.
The method can, in principle, be generalized to arbitrary two-body potentials that include attraction, for instance, by rescaling the maximum weight or oversampling.

\subsection{Hard-core repulsion}
\label{sec:sampleEV}

Here the focus is on a specific example of the interaction potential, namely the hard-core interaction potential given by
\eq{\label{eq:VHC}
V^{HC}(\bvar r_i, \bvar r_j) = 
\begin{cases}
  \infty, & |\bvar r_i - \bvar r_j| < \sigma~, \\
  0, & |\bvar r_i - \bvar r_j| \geq \sigma~.
\end{cases}   
}

With this choice of $V^{HC}(\bvar r_i, \bvar r_j)$, one deals with the system of hard spheres.
Here, $\sigma = 2 r_c$ is the hard-sphere diameter, and $r_c$ is the radius.
As discussed in Appendix~\ref{app:HS}, 
the equation of state of the hard-sphere system reduces to that of the excluded volume model in the dilute limit, provided that the excluded volume parameter $b$ of the latter model is taken as
\eq{\label{eq:bEV}
b = \frac{16 \pi r_c^3}{3}.
}
The hard-sphere and excluded volume model equations of state are very similar at $bn \lesssim 0.1$, where $n \equiv N/V$. Thus the hard-sphere model can mimic the excluded volume effect if this condition is met.
As discussed below, this is the case for the particlization stage in heavy-ion collisions.

For the case of hard-core potential~\eqref{eq:VHC} the probability density $\tilde P^{\rm coord}(\{\bvar r_i\})$ in Eq.~\eqref{eq:Pcoord} vanishes if any pair of particles overlap, i.e. if $|r_i - r_j| < \sigma$ for any $(i,j)$ pair.
In all other cases, $\tilde P^{\rm coord}(\{\bvar r_i\}) = 1$.

It follows that the system configuration can be sampled by the following algorithm involving a rejection sampling step:
\begin{enumerate}
    \item The coordinates of $N$ particles are sampled uniformly from the given (cubic) volume.
    \item If, for any $(i,j)$ pair of particles, they overlap, i.e., $|r_i - r_j| < \sigma$, the configuration is rejected, and one goes back to step 1.
    \item The momenta of the particles are sampled from the Maxwell-Boltzmann distribution.
\end{enumerate}
The procedure can be further sped up by combining steps 1 and 2: If a newly sampled particle overlaps with any previously sampled particles, the configuration can be rejected outright without needing to sample any remaining particles.

\subsection{Testing the sampling method}

\subsubsection{Canonical ensemble}

Here, sampling of the hard-sphere gas is performed, and particle number fluctuations in various coordinate space subsystems are analyzed.
The sampling is carried out for a fixed value of the hard-core radius $r_c$ and particle number density $n = N/V$, but for different values of the total number of particles $N$.
In particular, the scaled density $bn$ is fixed to $bn = 0.03$, where $b$ is given by Eq.~\eqref{eq:bEV}.
The value of $b$ is used to set the length scale, e.g. the dimensionless system volume reads $\tilde V = V/b = N / (bn)$, the cube length is $\tilde L = \tilde V^{1/3}$, and the dimensionless hard-core radius is $\tilde r_c = \tilde L \left[ \frac{3 (bn)}{16\pi N} \right]^{1/3} = \left(\frac{3}{16\pi}\right)^{1/3}$.
To minimize the effect of finite system size, periodic boundary conditions with minimum-image convention~(periodic box) are applied when checking the particles for overlap.
The results are also compared with the case where periodic boundary conditions are not applied~(single box).

After sampling the configurations, the first and second moments of particle number distributions inside various subvolumes of the coordinate space along the $z$ direction are studied. 
More specifically, the results are analyzed as a function of the total volume fraction $\alpha < 1$ covered by the subvolume.
A particle belongs to a subvolume $\alpha$ if its scaled coordinate $\tilde z$ is in the range $\frac{1-\alpha}{2} < \frac{\tilde z}{\tilde L} < \frac{1+\alpha}{2}$.

Monte Carlo simulations are performed for $N = 20,\,40,\,80,\,160$.
For the first moments, one observes that the mean number of particles in a subvolume is consistent within statistical errors with the relation $\mean{N}_\alpha = \alpha N$, that is, the particles are uniformly distributed throughout the volume on average, as they should.

For the second moment, the behavior of the scaled variance $\omega_\alpha \equiv \frac{\mean{N^2}_\alpha - \mean{N}_\alpha^2}{\mean{N}_\alpha}$ is studied.
Note that this quantity vanishes in the limit $\alpha \to 1$ since the total particle number is fixed.
For this reason, one looks at a scaled quantity $\tilde \omega_\alpha = \omega_\alpha / (1-\alpha)$.
In Ref.~\cite{Vovchenko:2020tsr} it was shown that this quantity is expected to approach the scaled variance in the grand-canonical limit, i.e., $\tilde \omega_\alpha \stackrel{N \to \infty}{\to} \omega_{\rm gce}$.
In the excluded volume model, one has~\cite{Gorenstein:2007ep}
\eq{
\omega_{\rm gce}^{\rm ev} = (1-bn)^2 \approx 0.941,
}
while in a more accurate Carnahan-Starling model~(see Appendix~\ref{app:HS}) one has $\omega_{\rm gce}^{\rm CS} \approx 0.942$.

\begin{figure}[t]
  \centering
  \includegraphics[width=.49\textwidth]{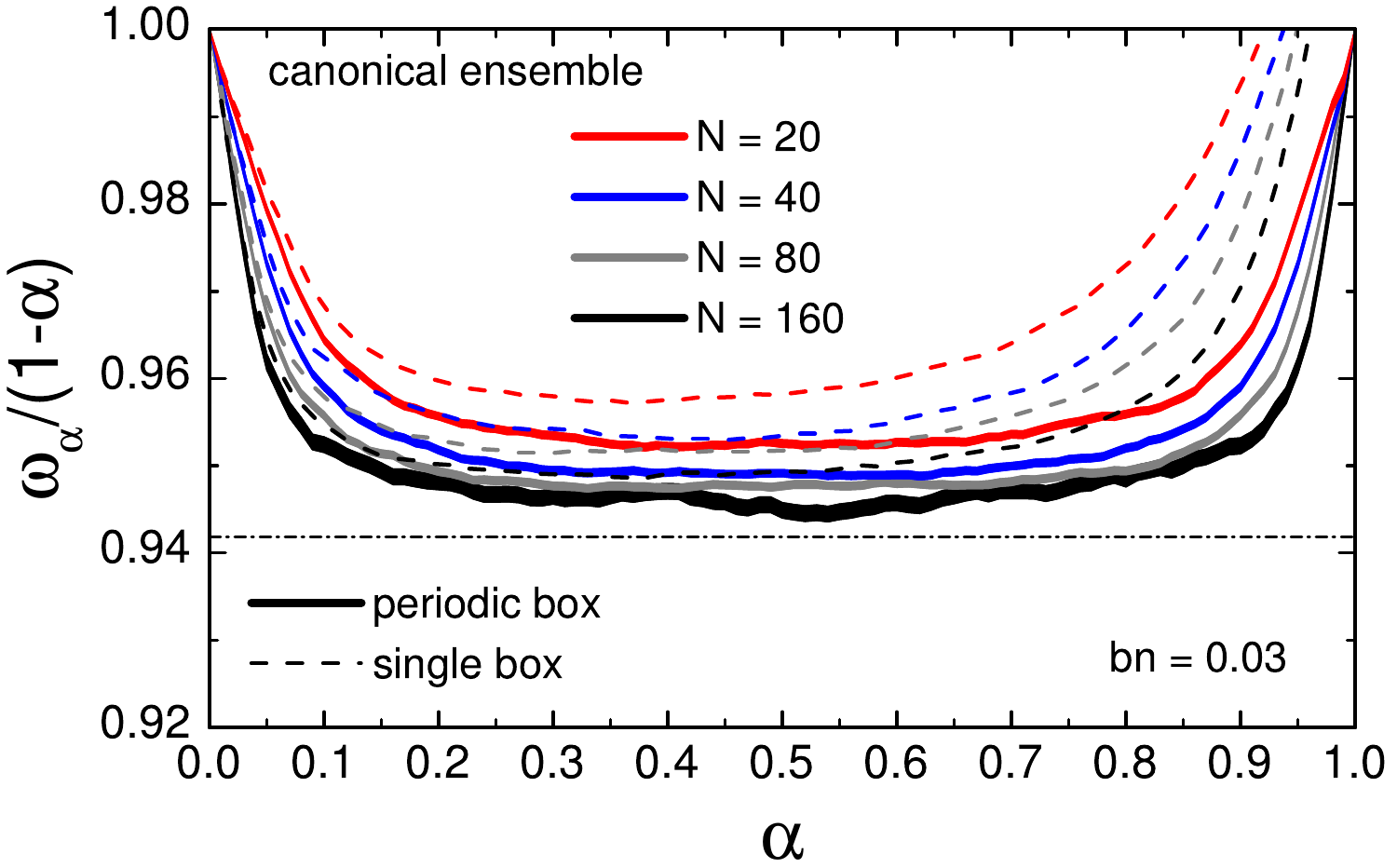}
  \caption{
   The dependence of the corrected scaled variance $\tilde \omega_\alpha = \omega_\alpha / (1-\alpha)$ on the acceptance fraction $\alpha$, evaluated in a box with~(bands) and without~(dashed lines) periodic boundary conditions resulting from the sampling of particles with a hard-core repulsion for different values of total particle number.
   The dash-dotted horizontal line corresponds to the expected thermodynamic limit of the hard-sphere equation of state calculated using the Carnahan-Starling model.
  }
  \label{fig:wCE}
\end{figure}

The $\alpha$-dependence of $\tilde w_\alpha$ resulting from the described sampling procedure is shown in Fig.~\ref{fig:wCE}, both with and without periodic boundary conditions.
The sample size is of the order of several million configurations in each case.
It is seen that the results show a suppression of scaled variance by magnitude, which is similar to the one given by the grand-canonical limit $\omega_{\rm gce}^{\rm CS}$, except for the $\alpha \to 0$ and $\alpha \to 1$ limits, where this quantity tends to the Poisson limit of unity.
Without periodic boundary conditions, the finite-size effects are more significant.

\subsubsection{Grand-canonical ensemble}

So far, the sampling has been discussed in the context of the canonical ensemble, i.e. for the case where the total number $N$ is fixed.
Consider the situation where the total number $ N $ itself fluctuates event-by-event.
One example of this scenario is the grand-canonical ensemble, where the system can exchange particles with a heat bath.
Even in the canonical ensemble, the total number of particles of a given type can fluctuate if the system is multi-component.
A relevant example for the present studies is the baryons and antibaryons in the canonical ensemble hadron resonance gas.
Even though the net number of baryons is precisely conserved, the individual numbers of baryons and antibaryons fluctuate.

The generic way to incorporate the fluctuations of $N$ is to fold the canonical ensemble procedure with the sampling of the total number $N$ in each event.
For example, the sampling of the grand-canonical excluded volume model is explored here.
Thus, simulations are performed not for a fixed total number $N$, but for a fixed mean number $\mean{N}$ that fluctuates following the grand-canonical ensemble distribution given by the EV model.
The sampling is performed in two steps:
\begin{enumerate}
    \item The total number $N$ is sampled from the grand-canonical excluded volume model, where the method described in Ref.~\cite{Vovchenko:2020kwg} is used.
    \item The coordinates of the $N$ particles are sampled using the method described in Sec.~\ref{sec:sampleEV}.
\end{enumerate}

\begin{figure}[t]
  \centering
  \includegraphics[width=.49\textwidth]{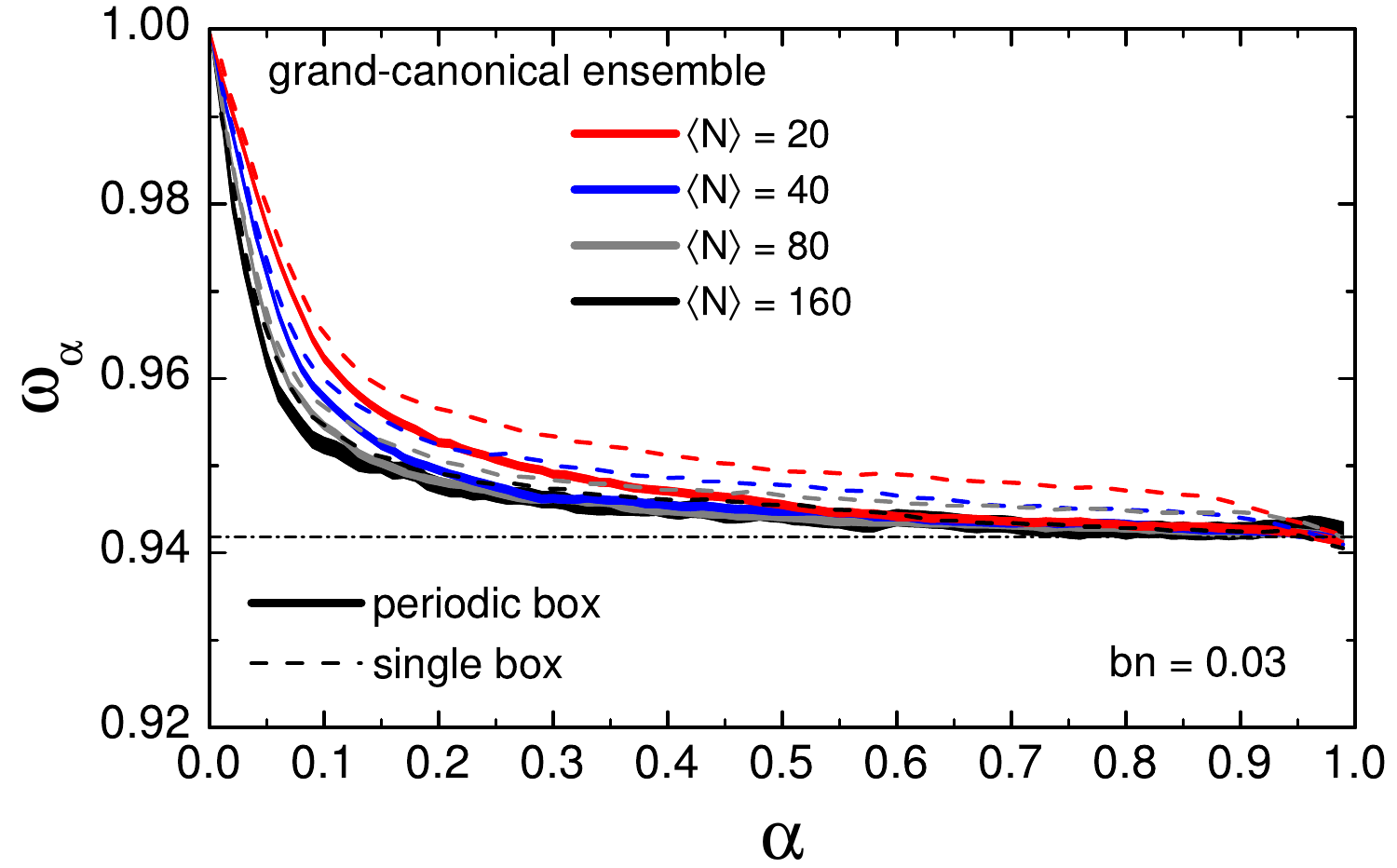}
  \caption{
   Same as Fig.~\ref{fig:wCE}, but for the uncorrected scaled variance $\omega_\alpha$ calculated within the grand-canonical ensemble. 
  }
  \label{fig:wGCE}
\end{figure}

Figure~\ref{fig:wGCE} shows the sampling results for $w_\alpha$ as functions of $\alpha$ in the grand-canonical ensemble.
For $\alpha \to 1$ the results approach the grand-canonical value, as expected.
The finite-size effects play a role at finite $\alpha$, keeping $\tilde w_\alpha$ slightly above the grand-canonical limiting value.
For $\alpha \to 0$, $\tilde w_\alpha$ approaches the Poisson limit of unity.

\subsection{SPR approximation}
\label{sec:SPR}

One can introduce the following approximation to the sampling procedure to achieve sufficiently fast sampling of an even larger number of particles.
Particles are sampled one by one. When the $i$th particle is sampled, its possible overlap with any of the already sampled particles is checked. If an overlap is detected, the sampled particle is rejected, and the process is repeated.
The difference from the exact method is that the already sampled $i-1$ particles are not rejected but retained.
This procedure is referred to as the single particle rejection (SPR) approximation, and it allows one to significantly improve the speed of the sampling procedure, especially for large systems.
The SPR approximation can be expected to be most accurate for moderate values of $bn$, i.e. for dilute systems.
Indeed, in dilute systems, it is improbable for a newly sampled particle to overlap with more than one other particle. Thus, there is no real need to reject all previously sampled particles if only a single overlap is detected.

\begin{figure}[t]
  \centering
  \includegraphics[width=.49\textwidth]{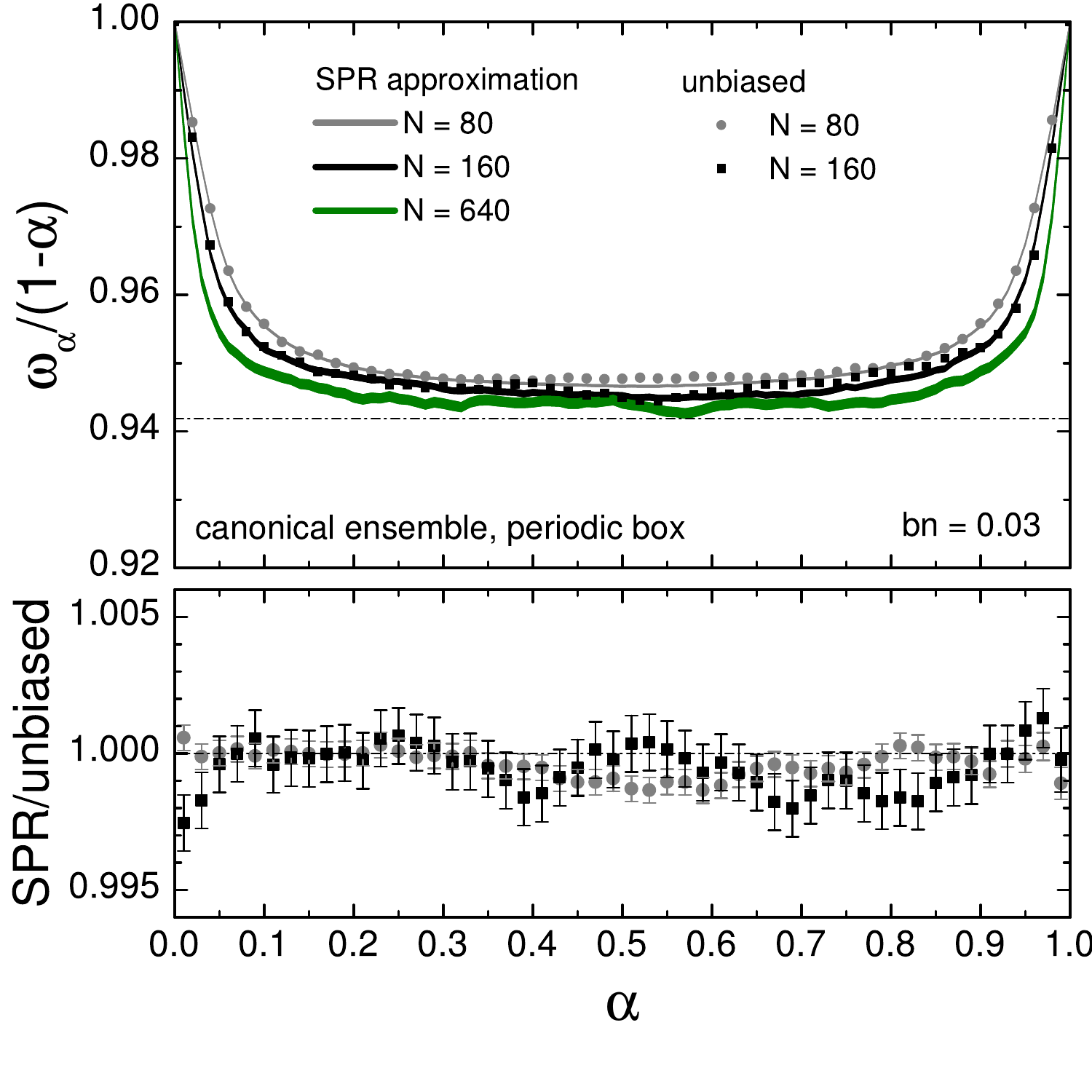}
  \caption{
   Same as Fig.~\ref{fig:wCE}, but calculated either within the SPR approximation~(bands) or without approximations~(symbols), for the periodic box only. The bottom panel shows the ratio of the approximate and unbiased calculations.
  }
  \label{fig:wCEperiodicfast}
\end{figure}

The accuracy of the SPR approximation can be tested with simulations. The results for $\tilde \omega_\alpha$ obtained using the SPR approximation are depicted in Fig.~\ref{fig:wCEperiodicfast}, for $N = 80,\,160,\,640$, for the periodic box only, and in the canonical ensemble.
The results for $N = 80$ and $160$ are consistent within errors with the results of the unbiased method, shown in Fig.~\ref{fig:wCEperiodicfast} by the symbols.
The results for $N = 640$ indicate the approaching of the grand-canonical value $\omega_{\rm gce}^{\rm ev}$ in the limit $N \to \infty$.

\begin{figure}[t]
  \centering
  \includegraphics[width=.49\textwidth]{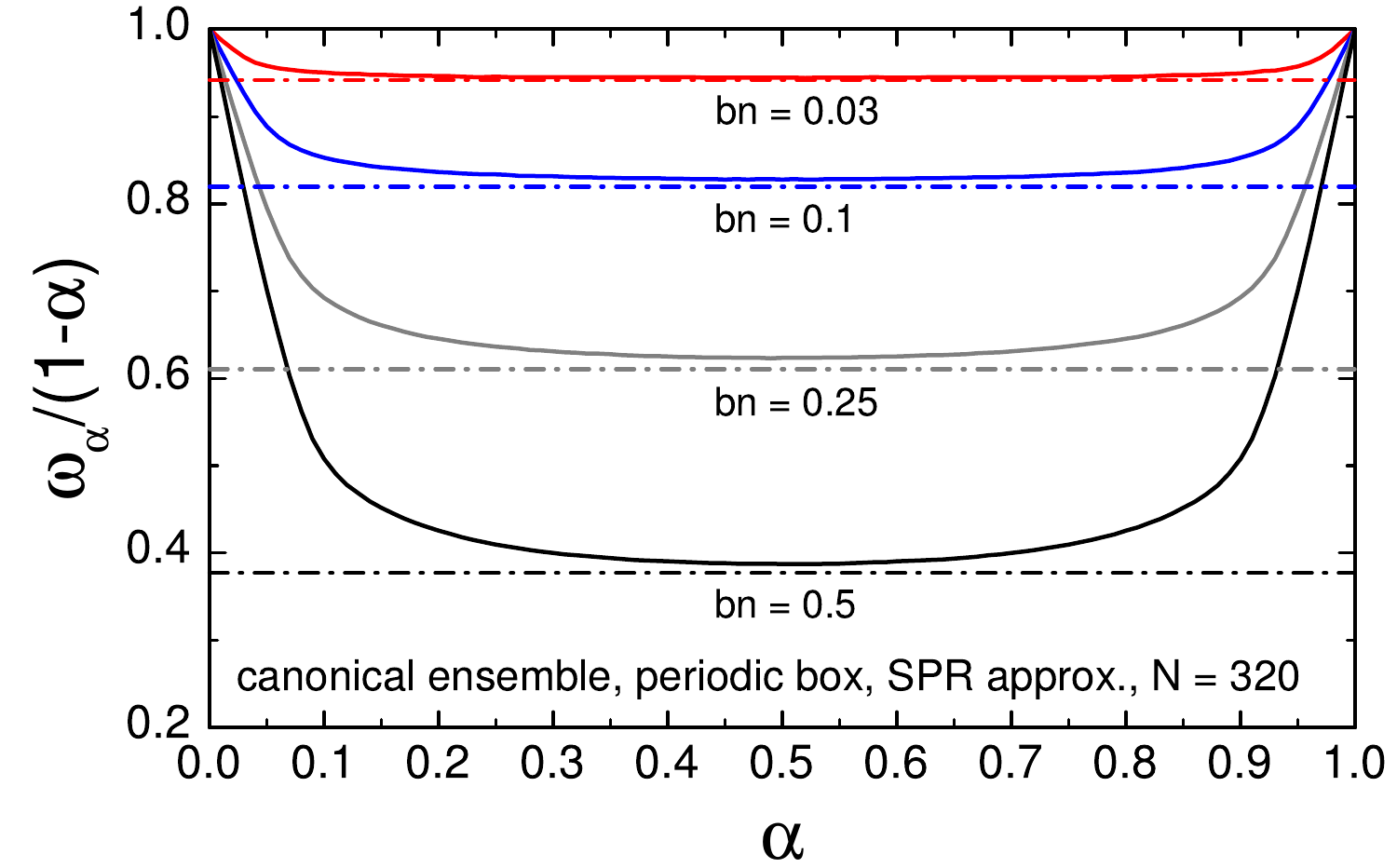}
  \caption{
   The dependence of the corrected scaled variance $\tilde \omega_\alpha = \omega_\alpha / (1-\alpha)$ on the acceptance fraction $\alpha$ calculated in a periodic box within the SPR approximation for different values of the scaled density $bn$. 
   The dash-dotted horizontal lines correspond to the expected thermodynamic limit of the hard-sphere equation of state calculated using the Carnahan-Starling model.
  }
  \label{fig:wCESPR}
\end{figure}

The method can be tested further by considering higher values of $bn$.
Figure~\ref{fig:wCESPR} depicts the results for $bn = 0.03,\,0.1,\,0.25,\,0.5$, all calculated for $N = 320$ within the SPR approximation.
The dash-dotted horizontal lines correspond to the grand-canonical values expected in the thermodynamic limit, as calculated within the Carnahan-Starling approximation~(see Appendix~\ref{app:HS}).
The Monte Carlo results consistently approach the grand-canonical values for all values of $bn$ considered.
One can conclude that the SPR approximation produces accurate results for the variance of particle number distributions at densities at least up to $bn = 0.5$.

Differences between the SPR approximation and the full sampler can be seen in more subtle observables than the cumulants.
An example based on the radial distribution function is discussed in Appendix~\ref{app:RDF}.

\section{Repulsive core in the Cooper-Frye hadron sampling procedure}
\label{sec:CooperFrye}

The system created in heavy-ion collisions is not a static box but a dynamically expanding, inhomogeneous fireball.
The momentum distribution for hadron species $j$ emerging from hydrodynamics is given by the Cooper-Frye formula~\cite{Cooper:1974mv}
\eq{\label{eq:CF}
\omega_p \frac{d N_j}{d^3 p} = \int_{\Sigma(x)} d \Sigma_\mu (x) \, p^\mu \, f_j[u^\mu(x) p_\mu;T(x),\mu_j(x)],
}
neglecting the shear and bulk viscous corrections.
Here $\Sigma(x)$ is the Cooper-Frye hypersurface, $d \Sigma_\mu (x)$ is the hypersurface element, $u^\mu(x)$ is the collective four-velocity, $p^\mu$ is the four-momentum and $f_j$ is the distribution function corresponding to a local thermodynamic equilibrium at a space-time point $x$.
The total mean number of particles can be obtained by integrating Eq.~\eqref{eq:CF} over the momenta.
This results in
\eq{\label{eq:meanTot}
\mean{N_j} & = \int_{\Sigma(x)} \mean{d N_j(x)}  \\
& =
\int_{\Sigma(x)} d \Sigma_\mu (x) u^\mu(x) \, n_j[T(x),\mu_j(x)]~.
}
Here $n_j$ is the equilibrium local rest frame density of particle species $j$ near the space-time point $x$. This density may generally contain the effect of hadronic interactions, such as excluded volume.
The total effective volume at particlization is
\eq{
V_{\rm eff} = \int_{\Sigma(x)} d \Sigma_\mu (x) u^\mu(x),
}
thus one can introduce the average effective density of particle species $j$ as $\mean{n_j} = \mean{N_j}/V_{\rm eff}$.

The Cooper-Frye sampling procedure that includes the effect of hard-core repulsion proceeds as follows:
\begin{enumerate}
    \item The total numbers of all the particle species are sampled. The effect of excluded volume on the distribution of the total particle numbers is included, where applicable, following the method described in Refs.~\cite{Vovchenko:2018cnf,Vovchenko:2020kwg} and using the average effective densities $\mean{n_j}$ as input.
    The exact global conservation laws are enforced through rejection sampling, where configurations that do not satisfy the exact conservation laws are discarded.
    
    \item The momenta and coordinates of each hadron are sampled one by one. To do that, first, the hypersurface element from which the given hadron is sampled is determined. This is done via the multinomial distribution where each volume element $x$ is weighted by $\mean{d N_j(x)}$, i.e. by the grand-canonical mean yield for the given hadron species emitted from that element. Then the momenta and coordinates of the hadron emitted from the chosen hypersurface element $x$ are sampled via the standard procedure.

    \item All pairs of hadrons with hard-core repulsion between them are checked for overlap. 
    If an overlap is detected, one goes back to the previous step.
    As in Sec.~\ref{sec:method}, to speed-up the procedure, one can check the overlap of a newly sampled particle with already sampled ones before sampling the remaining ones.
    As detailed below, each pair of hadron species can have a different value of the minimum distance $\sigma$ of their closest possible approach.
\end{enumerate}

Note that a numerical Cooper-Frye hypersurface might have negative volume elements with $d \Sigma_\mu (x) u^{\mu} (x) < 0$.
Here these volume elements are skipped, i.e., the presence of the $\theta$ function, $\operatorname{\theta}(d \Sigma_\mu (x) u^{\mu} (x))$, in all Cooper-Frye integrals is implied.

A few relevant details must be specified regarding the treatment of global conservation laws.
The total baryon number to be conserved is calculated from the 4$\pi$ mean hadron yields from Eq.~\eqref{eq:meanTot}.
The calculated value is then rounded to the nearest integer. To maintain consistency, all elements of the hypersurface $d \Sigma(x)$ are then rescaled by a common factor so that the baryon number calculated by Eq.~\eqref{eq:meanTot} coincides with the rounded integer number. This has only a minor effect, as the rescaling factor is very close to unity in all considered cases.
If strangeness is treated canonically, it is required to be equal to zero in each generated event.
Finally, the total net charge is constrained to reproduce the charge-to-baryon ratio of $Q/B = 0.4$. 
If the total electric charge satisfying the $Q/B = 0.4$ condition is not an integer, it is rounded to the nearest integer.

Note that the presented algorithm differs from most of the conventional methods used. 
There, hadrons are first sampled from each cell one by one, and then constraints from global conservation laws are imposed.
In this way, one may have to resort to rejecting all sampled particles if the global conservation laws are not satisfied.
The present algorithm deals with this problem more efficiently. Here, first, the total multiplicities of all hadrons are sampled, then the conservation laws are checked, and only after that does the sampling of hadrons' momenta and coordinates begin. The drawback is that, in the case of numerical Cooper-Frye hypersurfaces, this method requires storing the pre-computed multinomial probabilities for all hadron species and all Cooper-Frye hypersurface elements, which increases the memory requirements significantly. For example, the method requires about 5-15~Gb of RAM for sampling the central Au-Au collisions at RHIC-BES using Cooper-Frye hypersurfaces from the MUSIC code~\cite{MUSICinput}.
The present algorithm may also be less efficient for event-by-event hydrodynamics, as opposed to single-shot hydrodynamics used in the present study.

The procedure to check the overlap between two particles at the Cooper-Frye particlization has to be modified compared to the box case to account for the presence of collective motion, relativistic effects, and the fact that particles are emitted at different time moments.
This is achieved as follows:
\begin{itemize}
    \item First, both particles are boosted into their center-of-mass frame.
    \item Then, the particle emitted at the earlier time is propagated along the straight line to the time of the particle which was emitted later.
    \item Finally, the distance $|\bvar r_i- \bvar r_j|$ between the particles is calculated and checked if it is below the threshold value of 
    %$2 r^{s_i s_j}_c$, 
    $\sigma_{s_i s_j}$, where $s_i$ and $s_j$ is the species type of particles $i$ and $j$, respectively.
\end{itemize}

The hadron resonance gas at the Cooper-Frye particlization stage is a multi-component system.
Thus, in general, the hard-core repulsion between different hadron species is characterized not by a single excluded volume parameter $b$, but by a matrix $b_{\alpha \beta}$ of excluded volume parameters where each element corresponds to a distinct pair of hadron species. 
Note that this matrix need not be symmetric~\cite{Gorenstein:1999ce}.
The matrix $b_{\alpha \beta}$ can be used to determine the threshold distances $\sigma_{\alpha \beta}$.
As discussed in the framework of the multi-component excluded volume model~\cite{Gorenstein:1999ce}, the coefficients $b_{\alpha \beta}$ are related to $\sigma_{\alpha \beta}$ as 
\eq{
\frac{b_{\alpha \beta} + b_{\beta \alpha}}{2} = \frac{2 \pi \sigma_{\alpha \beta}^3}{3},
}
thus
\eq{\label{eq:sigab}
\sigma_{\alpha \beta} = \left( \frac{3 \overline{b}_{\alpha \beta}}{2\pi} \right)^{1/3},
}
where $\overline{b}_{\alpha \beta} = (b_{\alpha \beta} + b_{\beta \alpha})/2$.

The SPR approximation can be employed similarly to the box case in Sec.~\ref{sec:method}.
Using the SPR approximation is essential for sampling central collisions of heavy ions, where the system is so large that applying the full~(unbiased) method is prohibitively time-consuming.

It is instructive to summarize the list of approximations and possible limitations of the method:

\begin{itemize}
    \item[(i)] Total particle numbers $N_j$ are sampled from an auxiliary excluded volume HRG model characterized by constant volume $V_{\rm eff}$ and mean hadron yields $\mean{N_j}$, both calculated via the Cooper-Frye formula. The resulting excluded volume effect on the cumulants of $N_j$-distribution may be slightly different from the true result if the distribution of particle number densities across the Cooper-Frye hypersurface is inhomogeneous. The accuracy of this approximation for central Au-Au collisions at various beam energies is verified in the next section by comparing Monte Carlo results with analytic approximations, yielding only minor differences.

    \item[(ii)]  As the $N_j$-distribution is sampled from the excluded volume model rather than from the exact hard-sphere model distribution, for consistency, it is imperative that the excluded volume model provides an accurate approximation for the hard-spheres equation of state. As shown in the Appendix~\ref{app:HS}, this is satisfied with high precision for $bn \lesssim 0.10-0.15$.
    
    Both approximations (i) and (ii) concerning the $N_j$-distribution become less relevant when the canonical ensemble is applied, as the canonical ensemble effects then dominate the fluctuations of $N_j$.

    \item[(iii)] As the hard-core repulsion is an inherently non-relativistic concept, issues may arise when relativistic effects are strong. In particular, this can be an issue when the two overlapping particles have large relative velocities.
    This may happen, for instance, when the particles are emitted from different regions of the fireball characterized by significantly different collective velocities. However, as the hard-core repulsion is a short-range phenomenon, the issue becomes irrelevant in this case since these particles would not overlap in the coordinate space. 
    Particles that are close by in the coordinate space, on the other hand, would typically correspond to very similar collective velocities, thus their relative velocities are determined by the local temperature $T(x)$, which is usually of order 160~MeV or lower in heavy-ion collisions at particlization. 
    Therefore, the relativistic effects become less relevant for $m/T(x) \gg 1$, as is the case for baryons ($m_B \geq 938$~MeV/$c^2$), but could be important if hard-core repulsion is incorporated for lighter particles such as pions.
    
\end{itemize}

The Cooper-Frye sampling with short-range repulsion is implemented within the open source package~\texttt{Thermal-FIST}~\cite{Vovchenko:2019pjl} starting from version 1.4.

\section{Application of the method to proton cumulants in heavy-ion collisions}
\label{sec:HIC}

To illustrate the newly developed \texttt{FIST sampler} and the corresponding effect of hard-core repulsion, the behavior of (net-)proton cumulants in central collisions of heavy ions at various collision energies is studied.
The effect of (anti)baryon excluded volume on the cumulants has previously been studied at LHC~\cite{Vovchenko:2020gne} and RHIC-BES~\cite{Vovchenko:2021kxx} energies using different methods, thus, it is instructive to compare these results with the present method.

It is assumed here that excluded volume repulsion is present for all baryon-baryon and, by symmetry, all antibaryon-antibaryon pairs, as motivated by recent analyses of lattice QCD data on baryon number susceptibilities~\cite{Vovchenko:2017xad,Karthein:2021cmb}.
This implies $b_{\alpha \beta} = b > 0$ if either $\alpha, \beta \in B$ or $\alpha, \beta \in \bar{B}$, and $b_{\alpha \beta} = 0$ otherwise.
To be consistent with the earlier studies, the value of the baryon excluded volume parameter is chosen to be $b = 1$~fm$^3$, corresponding to a classical hard-core radius of around $r_c = 0.39$~fm.

\subsection{LHC}

\begin{figure*}[t]
  \centering
  \includegraphics[width=.49\textwidth]{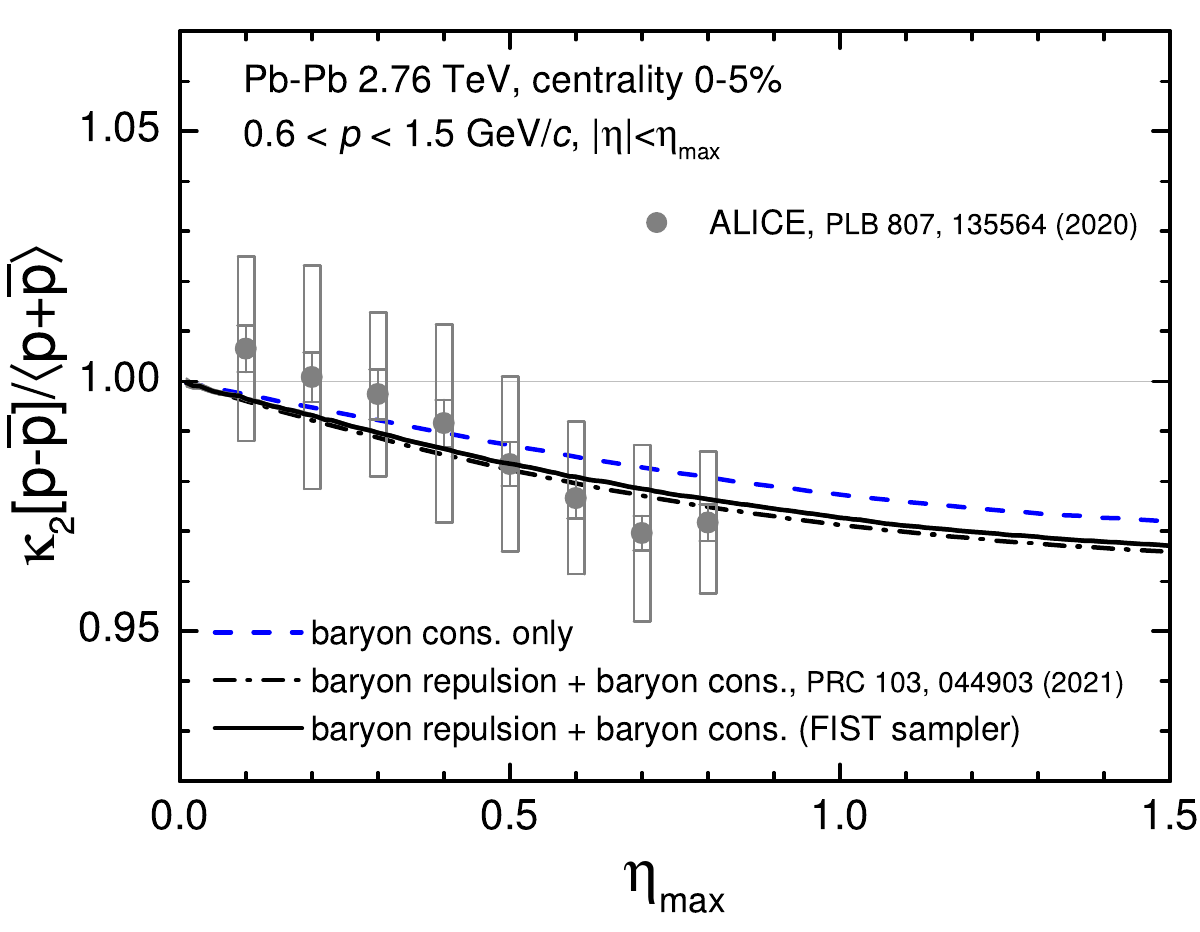}
  \includegraphics[width=.49\textwidth]{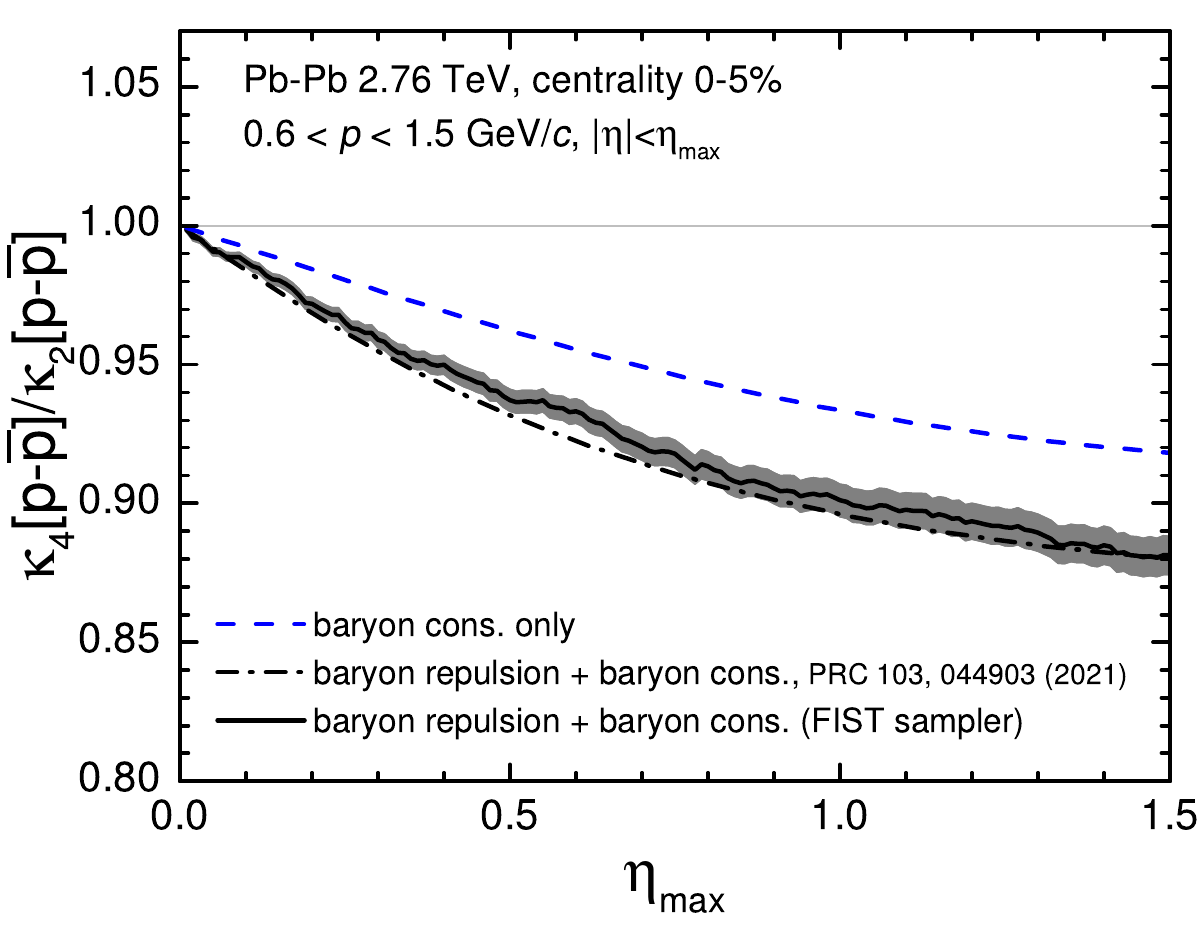}
  \caption{
  Cumulant ratios $\kappa_2[p-\bar{p}]/\mean{p+\bar{p}}$~(left panel) and $\kappa_4[p-\bar{p}]/\kappa_2[p-\bar{p}]$~(right panel) in 0-5\% central $\sNN = 2.76$~TeV Pb-Pb collisions as functions of the pseudorapidity cut $\eta_{\rm max}$.
  Calculations are performed within \fsampler~incorporating the effects of short-range baryon repulsion and global baryon conservation and shown by the black lines with bands.
  The dash-dotted black lines and dashed blue lines depict the calculations of Ref.~\cite{Vovchenko:2020kwg} within the subensemble sampler with and without the effect of baryon excluded volume, respectively.
  The symbols correspond to the experimental measurements of the ALICE Collaboration~\cite{Acharya:2019izy}.
  }
  \label{fig:ALICE}
\end{figure*}

First, the LHC energies are studied. 
More specifically, the 0-5\% central Pb-Pb collisions at $\sNN = 2.76$~TeV are analyzed, using a longitudinally boost-invariant Cooper-Frye hypersurface based on the blast-wave model.
The parametrization of the hypersurface is identical to a previous study~\cite{Vovchenko:2020kwg}, where the blast-wave model parameters are based on fits to the $p_T$ spectra within a single freeze-out scenario~\cite{Mazeliauskas:2019ifr}.
These parameters are uniform across the entire hypersurface.
The particlization temperature is $T = 160$~MeV, and the chemical potentials are vanishing.
For this choice of parameters, one has $b n_B = b n_{\bar{B}} \approx 0.029$, where $n_B$ and $n_{\bar{B}}$ are the number densities of baryons and antibaryons, respectively.

A total of $28.3$ million events is sampled. 
The events incorporate the baryon hard-core repulsion as described above, as well as the exact global conservation of baryon number.
The behavior of cumulants of net proton number distribution is analyzed. To minimize the statistical error, the fireball volume is reduced by factor 10, from $dV/dy = 400$~fm$^3$ to $dV/dy = 40$~fm$^3$. This reduction of the fireball volume is achieved through the reduction of its transverse radius from $R_\perp = 9$~fm to $R_\perp \approx 2.84$~fm. The analysis is therefore focused on the ratios of cumulants where the trivial dependence on the volume is canceled. 
For more details on this procedure to minimize the statistical error, see Ref.~\cite{Vovchenko:2020kwg}.

Figure~\ref{fig:ALICE} depicts the behavior of (a) the net-proton variance normalized over the Skellam distribution baseline, $\kappa_2[p-\bar p] / \mean{p + \bar p}$ and (b) the net-proton kurtosis ratio,  $\kappa_4[p-\bar p] / \kappa_2[p-\bar p]$.
The calculations are performed in the experimental acceptance used in the measurements performed by the ALICE Collaboration~\cite{ALICE:2019nbs}, corresponding to cuts in (anti)proton momentum, $0.6 < p < 1.5$~GeV/$c$, and pseudorapidity, $|\eta| < \eta_{\rm max}$.
The results in Fig.~\ref{fig:ALICE} are shown as a function of $\eta_{\rm max}$ by solid black lines with grey bands depicting the statistical uncertainties.
The results are compared with the earlier calculations of Ref.~\cite{Vovchenko:2020kwg} obtained in the framework of the so-called subensemble sampler~(dash-dotted black lines), as well as the ideal gas baseline that incorporates the effect of baryon conservation but not hard-core repulsion~(dotted blue lines).
Note that the calculations for $\kappa_4[p-\bar p] / \kappa_2[p-\bar p]$ do not account for the effect of volume fluctuations which would influence the measurements~\cite{Vovchenko:2020kwg}.

The \texttt{FIST sampler} results agree with the earlier results based on the subensemble sampler.
They also show visible deviations from the ideal gas baseline. This indicates that the \texttt{FIST sampler} provides a reasonable description of the excluded volume effect.
In contrast to the subensemble sampler, which is mainly restricted to longitudinally boost invariant systems such as those encountered at the LHC, the \texttt{FIST sampler} can also be straightforwardly applied to lower collision energies where boost invariance does not hold.
This is explored at the RHIC beam energy scan energies in the next section.

\subsection{RHIC-BES}

\begin{figure*}[t]
  \centering
  \includegraphics[width=.49\textwidth]{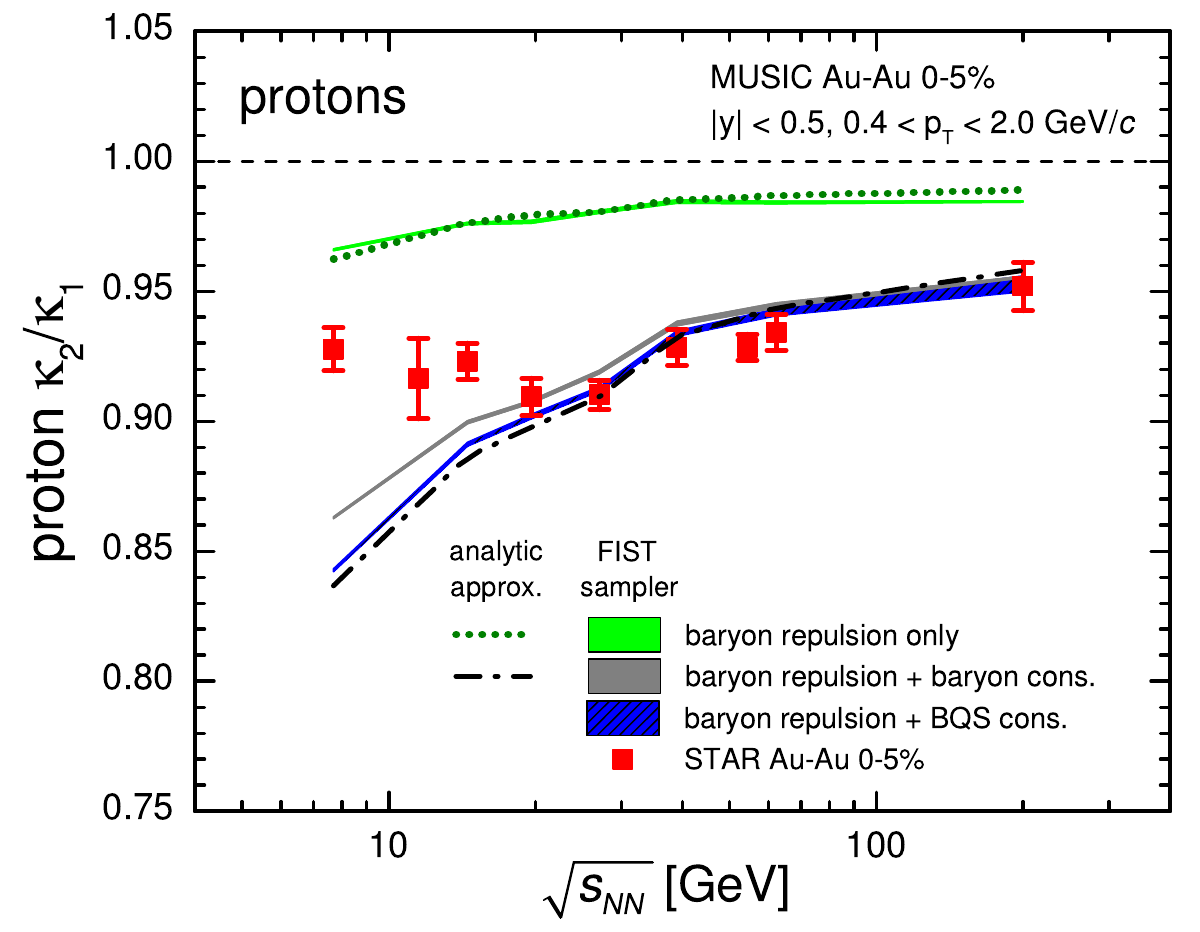}
  \includegraphics[width=.49\textwidth]{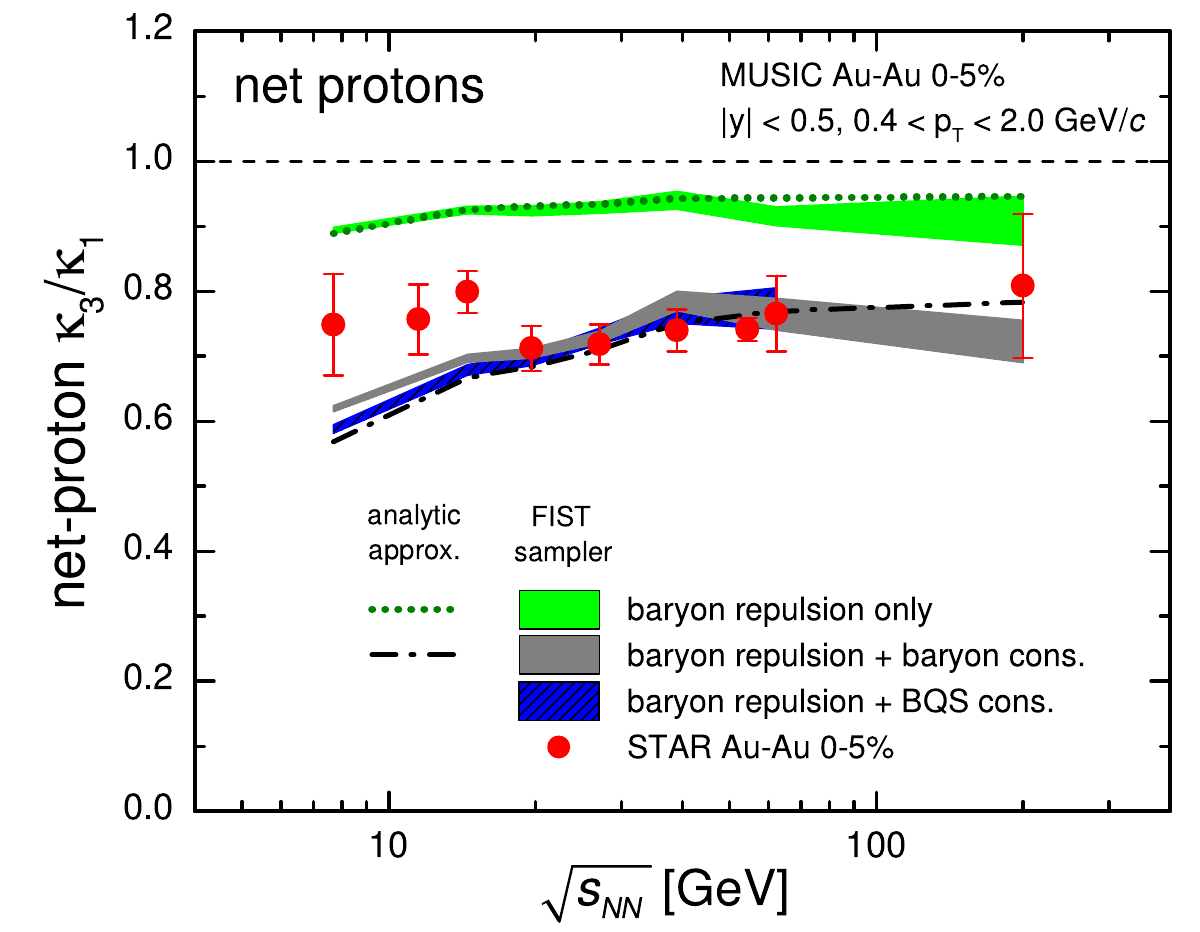}
  \caption{
  Beam energy dependence of the proton cumulant ratio $\kappa_2/\kappa_1$~(left panel) and net proton cumulant ratio $\kappa_3/\kappa_1$~(right panel) in 0-5\% central Au-Au collisions at RHIC-BES.
  The results depict \fsampler~calculations incorporating the baryon hard-core repulsion~(green bands), plus exact baryon conservation~(grey bands), plus exact conservation of electric charge and strangeness~(blue bands). The dotted green and dash-dotted black lines correspond to analytic calculations of Ref.~\cite{Vovchenko:2021kxx} with and without exact baryon conservation, respectively.
  The red symbols correspond to the experimental data of the STAR Collaboration~\cite{STAR:2021iop,STAR:2020tga}.
  }
  \label{fig:RHIC}
\end{figure*}

Net proton fluctuations have been measured by the STAR Collaboration in Au-Au collisions in collision energy range $\sNN = 7.7-200$~GeV~\cite{STAR:2020tga,STAR:2021iop}.
In Ref.~\cite{Vovchenko:2021kxx} the fluctuations have been analyzed based on relativistic hydrodynamics simulations with \texttt{MUSIC}~\cite{Shen:2020jwv}, incorporating effects of baryon conservation and excluded volume.
Calculations in Ref.~\cite{Vovchenko:2021kxx} were performed analytically, in two steps: (i) proton cumulants are calculated in the experimental acceptance using the Cooper-Frye formula in the grand-canonical limit, including the excluded volume effect; (ii) correction for baryon number conservation is performed using a method called SAM-2.0~\cite{Vovchenko:2021yen}.

Here, the calculations of proton cumulants are performed using the \texttt{FIST sampler}, by sampling 0-5\% central Au-Au collisions at energies $\sNN = 7.7,\,14.5,\,19.6,\,27,\,39,\,62.4$ and $200$ GeV.
The sampling is performed with and without the effect of exact baryon conservation, and it uses the same Cooper-Frye hypersurfaces produced by the \texttt{MUSIC} code that is based on \cite{Shen:2020jwv} and available at~\cite{MUSICinput}.
For each energy, at least several million events have been generated.
These calculations are compared to the analytic approximations of Ref.~\cite{Vovchenko:2021kxx} in Fig.~\ref{fig:RHIC}, for (a) the second proton cumulant ratio $\kappa_2/\kappa_1$ and (b) the third order net proton cumulant ratio $\kappa_3/\kappa_1$.
The \fsampler~results are shown by the bands, while the lines show the analytic results.

In the absence of baryon conservation~(\emph{baryon repulsion only}, green lines and bands in Fig.~\ref{fig:RHIC}) deviations of the cumulant ratios from the baseline of unity are driven solely by baryon repulsion, which leads to a modest suppression. The \fsampler~results agree with the analytic approximations, validating the method at RHIC-BES energies.

Incorporating exact baryon conservation~(\emph{baryon repulsion + baryon cons.}, black lines, and grey bands in Fig.~\ref{fig:RHIC}) leads to a more potent suppression of the cumulant ratios.
In this case, the analytic and Monte Carlo results agree at high collision energies~($\sNN \gtrsim 40$~GeV), but deviations between the two are visible at lower $\sNN$. 
These deviations can be attributed to inaccuracies of the SAM-2.0 framework at lower collision energies. As discussed in the original publication~\cite{Vovchenko:2021yen}, it would tend to overestimate the effect of (repulsive) interactions.
Comparison with the \fsampler~allows one to quantify the accuracy of this approximation at RHIC-BES conditions, revealing a modest overestimation of excluded volume effect on proton cumulants which gets worse as collision energy is decreased.

As a new application of the \fsampler, one can simultaneously incorporate the effects of baryon repulsion and exact conservation of all three QCD conserved charges: baryon number, electric charge, and strangeness.
The electric charge conservation, in particular, can influence the fluctuations of the proton number since the proton carries the electric charge.
At high collision energies, the dominant electric charge carriers are pions. Thus, the effect of electric charge conservation on protons is expected to be small~\cite{Vovchenko:2020gne}.
As the collision energy decreases, the fraction of charge carried by protons increases, and so does the effect of charge conservation on proton cumulants.
This is reflected in the \fsampler~results shown in Fig.~\ref{fig:RHIC} by the blue bands, indicating additional suppression of (net-)proton $\kappa_2/\kappa_1$ and $\kappa_3/\kappa_1$ due to electric charge conservation, which becomes visible at $\sNN \lesssim 40$~GeV.
Interestingly, at all collision energies considered, the \fsampler~results with exact conservation of baryon number, electric charge, and strangeness are in good agreement with the analytical calculations of proton cumulants from Ref.~\cite{Vovchenko:2021yen} that incorporate baryon conservation but not electric charge and strangeness.
Thus, it appears that the additional suppression due to electric charge conservation at all energies is of the same magnitude as the overestimation of the excluded volume effect in the analytical calculation of Ref.~\cite{Vovchenko:2021yen} with SAM-2.0.
Although this coincidence appears to be purely accidental, it indicates that the results of Ref.~\cite{Vovchenko:2021yen} serve as an accurate baseline for proton cumulants incorporating non-critical effects like baryon repulsion and exact conservation of multiple conserved charges.

\subsection{GSI-SIS}

\begin{figure*}[t]
  \centering
  \includegraphics[width=.32\textwidth]{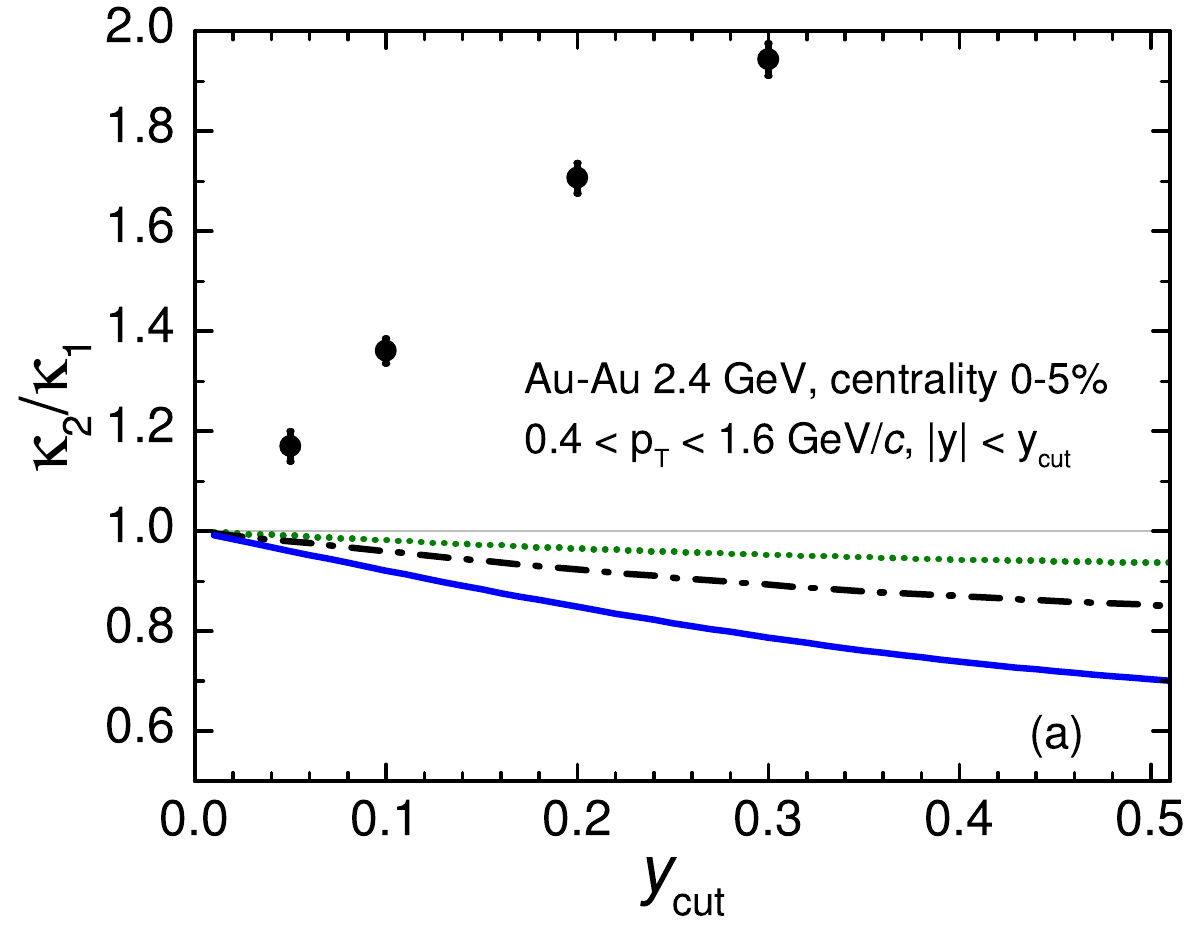}
  \includegraphics[width=.32\textwidth]{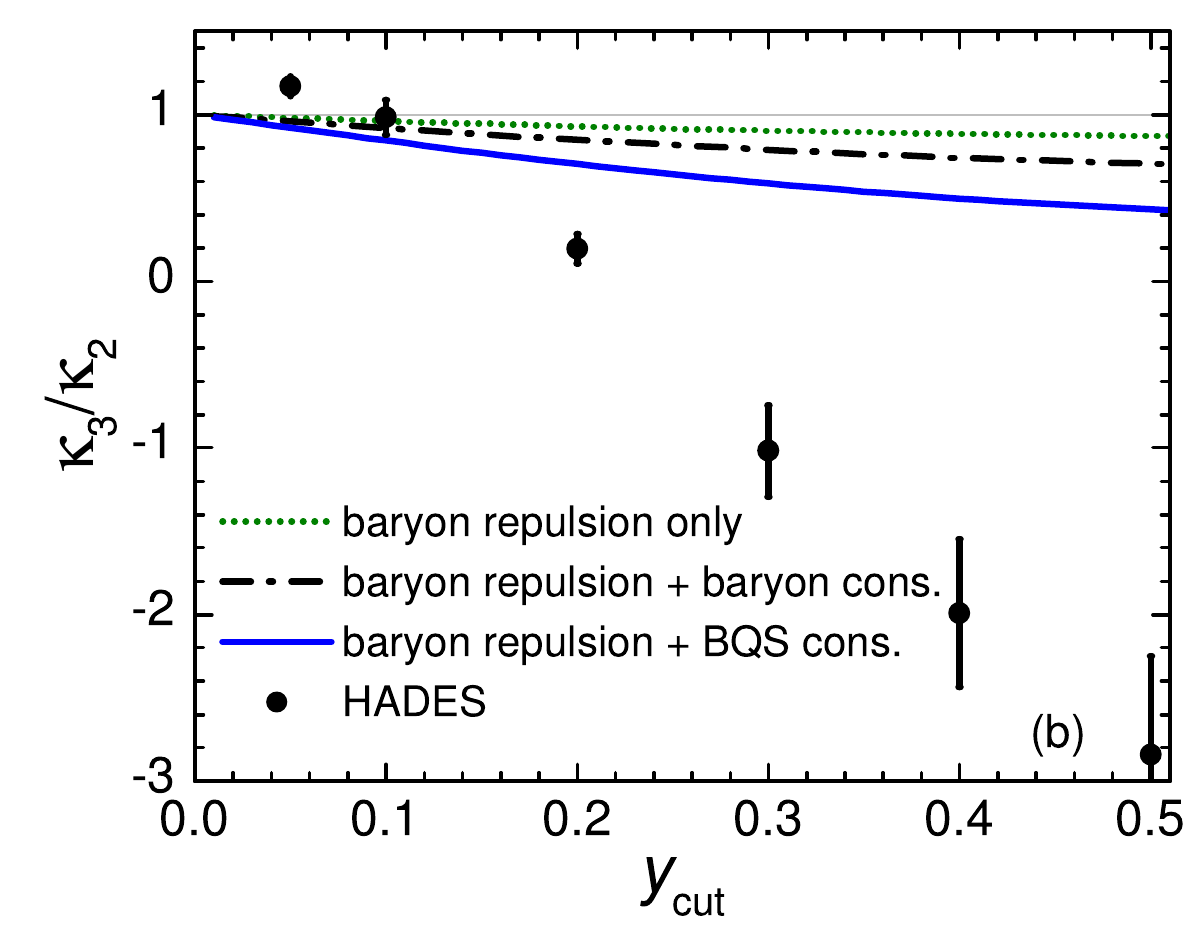}
  \includegraphics[width=.32\textwidth]{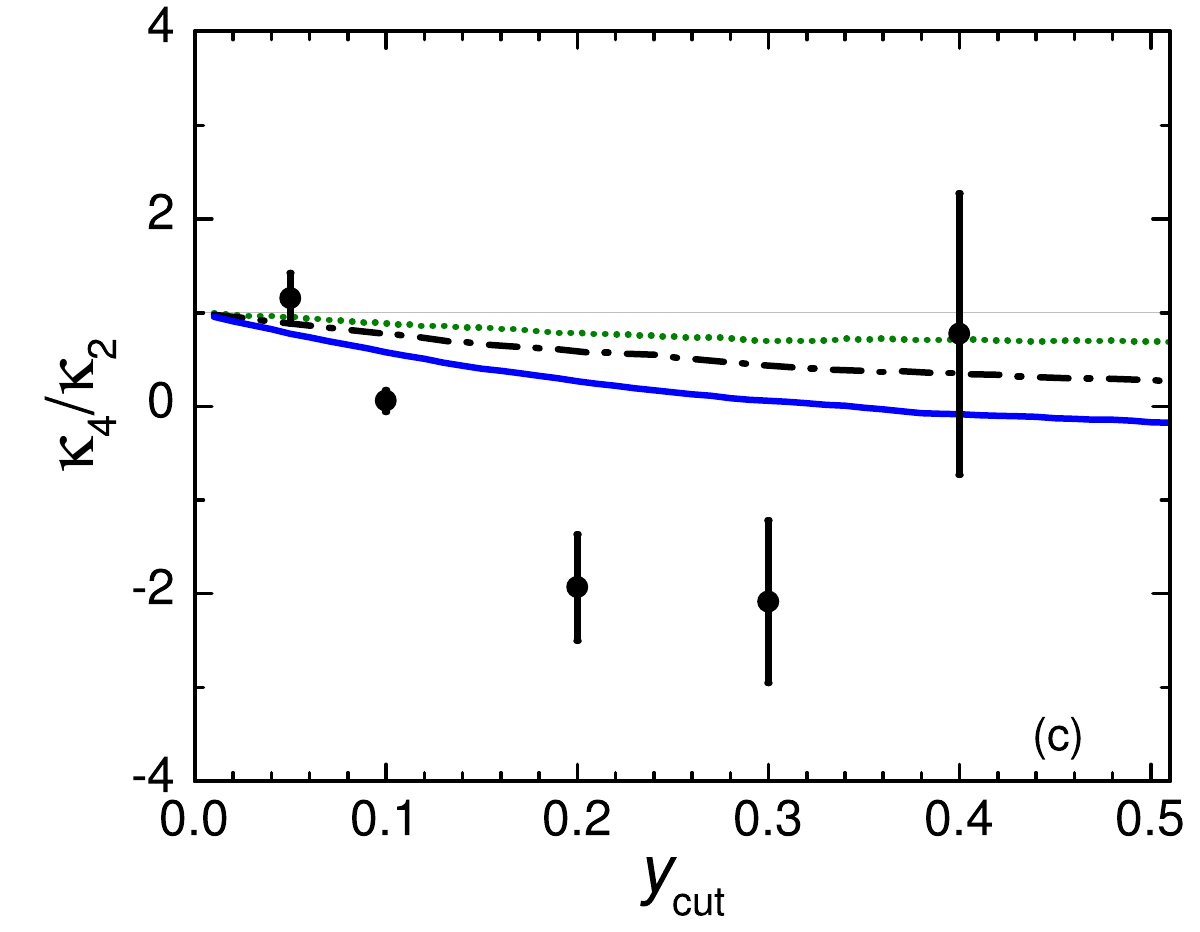}
  \caption{
  Cumulant ratios (a) $\kappa_2/\kappa_1$, (b) $\kappa_3/\kappa_2$, and (c) $\kappa_4/\kappa_2$ of the proton number distribution in 0-5\% central $\sNN = 2.4$~GeV Au-Au collisions in the experimental acceptance of the HADES Collaboration, as calculated within \fsampler.
  The results are shown as a function of the rapidity cut $y_{\rm cut}$. 
  The calculations incorporate the baryon hard-core repulsion~(dotted green lines), plus exact baryon conservation~(dash-dotted black lines), plus exact conservation of electric charge and strangeness~(solid blue lines).
  The black symbols correspond to the experimental data of the HADES Collaboration~\cite{HADES:2020wpc}.
  }
  \label{fig:HADES}
\end{figure*}

The final application of the new method concerns the Au-Au collisions at $\sNN = 2.4$~GeV, as probed by the HADES experiment at GSI-SIS.
The experimental data~\cite{HADES:2020wpc} indicate sizable multiproton correlations.
These were recently analyzed in the framework of a fireball model~\cite{Vovchenko:2022szk}.
The Cooper-Frye hypersurface was parameterized in the framework of the Siemens-Rasmussen model with a Hubble-like collective flow, which was found to provide a reasonable description of the $p_T$ spectra of pions and protons~\cite{Harabasz:2020sei}, while the temperature $T \approx 70$~MeV and baryochemical potential $\mu_B \approx 875$~MeV were extracted from hadron yields~\cite{Motornenko:2021nds}.
Here this parametrization of the hypersurface is used in \fsampler~to study the behavior of proton cumulants influenced by conservation laws and baryon repulsion.

At $\sNN = 2.4$~GeV, a significant fraction of protons is bound into light nuclei in the final state. Based on the preliminary HADES data~\cite{Harabasz:2020sei}, one can estimate that about $37.5$\% of the final-state protons are bound. The light nuclei are not incorporated into the \fsampler~employed in the present work. Thus, to model this effect, following Ref.~\cite{Vovchenko:2022szk}, each proton emitted from the fireball is assumed to be bound into light nuclei with a probability of $q_{\rm nucl} = 0.375$ and not contribute to the measured cumulants of the proton number.

The corresponding results for proton number cumulant ratios $\kappa_2/\kappa_1$, $\kappa_3/\kappa_2$, and $\kappa_4/\kappa_2$ are shown in Fig.~\ref{fig:HADES} as functions of the rapidity cut $y_{\rm cut}$ in the experimental acceptance of the HADES experiment.

The results show a similar pattern to RHIC-BES energies: both the baryon conservation and repulsion suppress the cumulant ratios, with baryon conservation having a stronger effect. The additional effect of electric charge conservation~(blue lines) is even more notable, which suppresses the cumulant ratios even further. The combined effect of baryon and electric charge conservation is more significant than that of baryon repulsion.
Note that both the total baryon and electric charge of the participant matter fluctuates event by event in the experiment. This is in contrast to the \fsampler~where they are both fixed. The experimental data of the HADES collaboration~\cite{HADES:2020wpc}, shown in Fig.~\ref{fig:HADES} by black symbols, is corrected for volume~(participant) number fluctuations, which justifies the canonical treatment of baryon number for the participant matter. However, the total electric charge of the participants still fluctuates; thus, the \fsampler~results in Fig.~\ref{fig:HADES} should be considered an upper bound on the effect of exact conservation of the electric charge on the proton number cumulants.
It is also clear that the model fails to describe the experimental data even qualitatively. For example, the scaled variance $\kappa_2/\kappa_1$ increases with the rapidity cut $y_{\rm cut}$ in the experiment, as opposed to the suppression predicted due to conservation laws and short-range repulsion. This indicates that the behavior of the proton cumulants in the experimental data is not driven by the conservation laws or short-range repulsion that are incorporated in \fsampler.
For a discussion of the various possibilities, including the QCD critical point, see Ref.~\cite{Vovchenko:2022szk}

The present analysis indicates that the exact conservation of multiple conserved charges, as opposed to only that of the baryon number, is essential for quantitative analysis of proton number cumulants at moderate collision energies, $\sNN \lesssim 7.7$~GeV.
The \fsampler~allows one to evaluate the corresponding baselines in this energy regime, which can then be used to analyze future experimental data coming from the fixed-target program at RHIC~\cite{STAR:2021fge} or the CBM experiment at FAIR~\cite{CBM:2016kpk}.

For comparisons of the \fsampler~results in the SIS-HADES regime with analytic approximations, see Appendix~\ref{app:SAMHADES}.

\section{Conclusions and outlook}
\label{sec:summary}

This study introduced short-range repulsive correlations into the Cooper-Frye hadron sampling procedure.
This has been achieved through a rejection sampling step that prohibits any two particles with repulsive interactions from overlapping in the coordinate space, effectively modeling the excluded volume phenomenon.
The effect introduces negative correlations between particles visible in normalized cumulants of their distributions.
The new method -- called the \fsampler~-- incorporates this effect and simultaneously allows for both the canonical and grand-canonical treatments of the QCD conserved charges.

\fsampler~was validated in a periodic box setup, where the method yields the behavior of the scaled variance of particle number, which is consistent with the equation of state of hard spheres.
\fsampler~was then used to model the excluded volume effect in (anti)baryon-(anti)baryon interaction at the particlization stage of heavy-ion collisions in a broad collision energy range, $\sNN = 2.4-2760$~GeV.
In a longitudinally boost-invariant scenario at LHC energies, the new method produces the behavior of net proton cumulants up to the fourth order, which is consistent with an earlier study of~\cite{Vovchenko:2020kwg} that used the so-called subensemble sampler. 

The advantage of \fsampler~becomes evident at lower collision energies, where boost invariance no longer holds and where the application of the subensemble sampler is challenging.
The new method allowed one to incorporate the simultaneous effects of baryon repulsion and baryon, electric charge, and strangeness conservation at RHIC-BES and SIS-GSI energies. 
In particular, the effect of electric charge conservation on proton number cumulants becomes increasingly important as the collision energy is decreased, $\sNN \lesssim 40$~GeV, due to a smaller number of pions relative to protons at lower energies.
The resulting cumulant ratios of the (net) proton distribution are consistent with estimates of Ref.~\cite{Vovchenko:2021kxx} that were obtained by employing analytic approximations for baryon conservation and repulsion and neglecting electric charge conservation.
The reason for this coincidence at lower energies appears to be that the analytic method of Ref.~\cite{Vovchenko:2021kxx} slightly overestimates the effect of baryon repulsion in the canonical ensemble, which at lower collision energies is approximately compensated for by the electric charge conservation in \fsampler.
Due to this coincidence, the results of Ref.~\cite{Vovchenko:2021kxx} can be taken as a baseline for (net) proton cumulants at RHIC-BES that incorporates non-critical effects such as excluded volume and multiple global conservation laws.
The mechanisms above do not describe the experimental data of the HADES Collaboration~(Fig.~\ref{fig:HADES}), indicating that proton number cumulants are driven by different mechanisms that still need to be clarified.

\fsampler~has several future potential applications. For instance, one can couple its output with a hadronic afterburner such as UrQMD~\cite{Bass:1998ca,Bleicher:1999xi} or SMASH~\cite{Weil:2016zrk}, which should ideally include the short-range repulsion for consistency~\cite{Sorensen:2020ygf}, and study the effect of the hadronic phase on proton number cumulants, in particular, that of baryon annihilation~\cite{Savchuk:2021aog,Garcia-Montero:2021haa}.
Although the effects of hadronic afterburner have not been studied in the present work, the \fsampler~ code~\cite{FSgithub} readily supports output tailored for use with UrQMD afterburner~\cite{urqmd-afterburner-toolkit}.
Since the method offers the full flexibility of an event generator, observables other than proton number cumulants, such as, for example, balance functions~\cite{Bass:2000az} or distributions of other hadronic species~\cite{STAR:2019ans} can be studied as well.
In particular, one may revisit the Bayesian constraints on transport coefficients of QCD matter and the switching temperature for the transition from hydrodynamics to transport~\cite{JETSCAPE:2020shq,JETSCAPE:2020mzn} by employing the new particlization routine presented here in such an analysis.
One can also straightforwardly incorporate the possible flavor dependence of baryon repulsion suggested by recent analyses of lattice QCD data on off-diagonal susceptibilities~\cite{Motornenko:2020yme,Karthein:2021cmb,Bollweg:2021vqf} by an appropriate choice of the corresponding threshold distances $\sigma_{\alpha \beta}$ in Eq.~\eqref{eq:sigab}.
It can also be interesting to see how the presence of baryonic hard-core repulsion might affect the light nuclei production, such as employed in phase-space coalescence prescriptions~\cite{Sombun:2018yqh,Hillmann:2021zgj}.
These applications will be the subject of future studies.

Using the single particle rejection (SPR) approximation was essential to sample central collisions of heavy ions in practical applications of the new routine, as applying a direct unbiased method was prohibitively time-consuming. The implication is that the sampler is not fully unbiased, even though the tests in a box setup indicate that the sampling of particle number cumulants remains accurate.
It is thus advisable to consider other techniques, such as oversampling, to produce a fast, unbiased sampler.

The sampling with short-range repulsion described in this work has been implemented into the Monte Carlo event generator within the open source package~\texttt{Thermal-FIST}~\cite{Vovchenko:2019pjl}, starting from version 1.4.
The code for the sampling of various central collisions of heavy ions discussed in the present work is available at~\cite{FSgithub}.

\begin{acknowledgments}

%\section*{Acknowledgements}
%\emph{Acknowledgments.} 
The author is grateful to Mark Gorenstein for reading the manuscript and for useful comments.
The author thanks Volker Koch and Chun Shen for fruitful discussions and collaboration on related projects and Dylan Neff for stimulating discussions.
The author also acknowledges the hospitality of the staff at the Frankfurt Institute for Advanced Studies where part of this work was done.
This work was supported through the U.S. Department of Energy, 
Office of Science, Office of Nuclear Physics, under contract number 
DE-FG02-00ER41132. 
Computational resources were provided by the Frankfurt Center for Scientific Computing (Goethe-HLR).

\end{acknowledgments}

\appendix

\section{The equation of state and particle number fluctuations in the system of hard spheres}
\label{app:HS}

The equation of state that describes the system of classical particles that interact through a hard-core potential~\eqref{eq:VHC} can be written in terms of the compressibility factor $Z \equiv P / (nT)$:
\eq{
\frac{P}{nT} = g(\eta),
}
where $\eta = \frac{bn}{4} = \frac{4\pi r_c^3}{3} \frac{N}{V}$ is the so-called packing fraction.
The explicit expression for the function $g(\eta)$ is not known, but different approximations are available that model the equation of state accurately when the values of $\eta$ are not too large. One well-known approximation is the van der Waals excluded volume model where one has $P^{\rm ev} = \frac{nT}{1-bn}$, thus
\eq{\label{eq:Pev}
g^{\rm ev}(\eta) = (1-4\eta)^{-1}.
}
A more involved analytical approximation of the equation of state of hard spheres is given by the Carnahan-Starling model~\cite{carnahan1969equation}:
\eq{\label{eq:Pcs}
g^{\rm CS}(\eta) = \frac{1+\eta+\eta^2-\eta^3}{(1-\eta)^3}.
}

\begin{figure}[t]
  \centering
  \includegraphics[width=.49\textwidth]{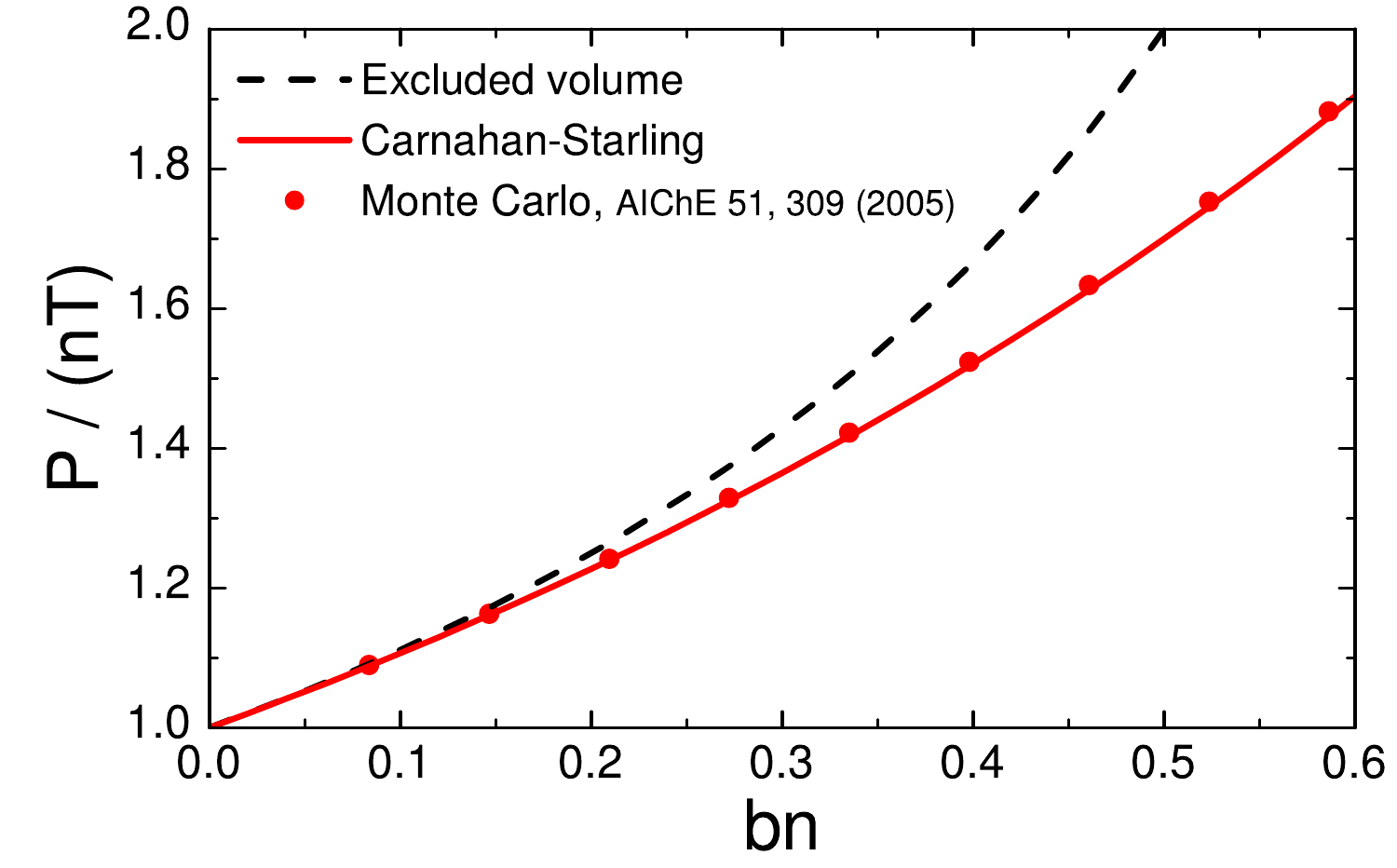}
  \caption{
   Density dependence of the compressibility factor $Z = P/(nT)$ in the system of hard spheres, as calculated in the excluded volume~(dashed black line) and Carnahan-Starling~(solid red line) approximations.
   The symbols show the numerical Monte Carlo results from Ref.~\cite{wu2005hard}.
  }
  \label{fig:HS.EoS}
\end{figure}

The accuracy of the excluded volume and Carnahan-Starling models can be verified by comparing the results to numerical Monte Carlo simulations of the hard-sphere system.
This comparison is depicted in Fig.~\ref{fig:HS.EoS}, where the dependence of the compressibility factor $Z$ on $bn$, calculated in the two models, is compared to the numerical simulations of~\cite{wu2005hard}.
The excluded volume model accurately describes the compressibility factor up to $bn \approx 0.10-0.15$, while it overestimates the numerical data for $Z$ at higher densities.
On the other hand, the Carnahan-Starling model accurately describes the Monte Carlo data for all the densities considered, that is, up to at least $bn = 0.6$.

\begin{figure}[t]
  \centering
  \includegraphics[width=.49\textwidth]{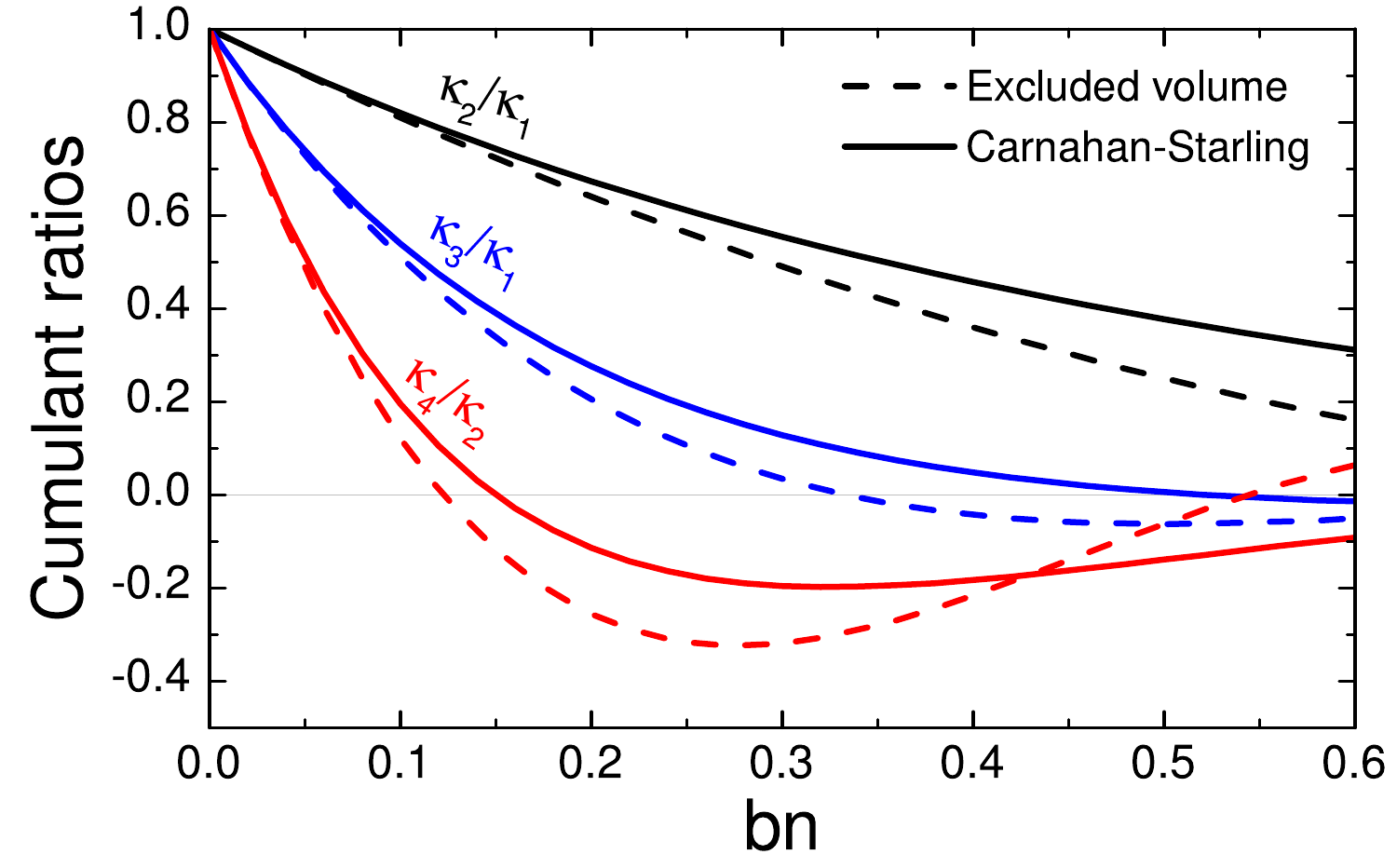}
  \caption{
   Density dependence of cumulant ratios $\kappa_2/\kappa_1$~(black, top), $\kappa_3/
   \kappa_1$~(blue, middle), $\kappa_4/\kappa_2$~(red, bottom) of the grand-canonical particle number distribution calculated for the hard-sphere system within the excluded volume~(dashed lines) and Carnahan-Starling~(solid lines) approximations.
  }
  \label{fig:HS.kappa}
\end{figure}

Cumulants of the grand-canonical particle number distribution are given by
\eq{
\kappa_n = V \, T^{n-1} \, \left(\frac{\partial^n P}{\partial \mu^n}\right)_T.
}
For $n = 1$ one has $\kappa_1 \equiv \mean{N}$, therefore, for $n > 1$ the cumulants read
\eq{
\kappa_n = T^{n - 1} \, \left(\frac{\partial^{n-1} \mean{N}}{\partial \mu^{n-1}}\right)_T.
}
Using the thermodynamic identity $\left(\frac{\partial{\mean{N}}}{\partial \mu}\right)_{T,V} = \mean{N} / \left(\frac{\partial{P}}{\partial n}\right)_{T}$, one can evaluate the derivatives $\left(\frac{\partial^{n-1} \mean{N}}{\partial \mu^{n-1}}\right)_T$ iteratively, provided that the pressure $P$ is known as a function of density $n$.
This is the case for the excluded volume~[Eq.~\eqref{eq:Pev}] and Carnahan-Starling~[Eq.~\eqref{eq:Pcs}] models.

Figure~\ref{fig:HS.kappa} depicts the behavior of the grand-canonical cumulant ratios $\kappa_2/\kappa_1$, $\kappa_3/\kappa_1$, and $\kappa_4/\kappa_2$ as a function of the scaled density $bn$.
Given the accuracy of the Carnahan-Starling approximation for the compressibility $Z$~(Fig.~\ref{fig:HS.EoS}), it is expected that the corresponding calculations of the cumulant ratios accurately reflect the grand-canonical particle number distribution in the hard-sphere system in the considered density range.
Deviations of the excluded volume model from the Carnahan-Starling model become more visible in higher-order cumulants.
The excluded volume model provides accurate results up to $bn \approx 0.10$.

\section{Radial distribution function}
\label{app:RDF}

The radial distribution function $g(r)$ describes how the density of particles varies around a reference particle at $r = 0$ relative to the expectation based on the mean particle number density $n = N / V$.
More specifically, $g(r)$ is defined so that the local particle number density at a distance $r$ from the reference particle is equal to $n \, g(r)$.
Deviations of $g(r)$ from unity signal the presence of correlations between particles.

The radial distribution function for the hard-sphere system has been studied for a long time~\cite{PhysRevLett.10.321,doi:10.1063/1.1734272}.
As the particles cannot overlap, one has $g(r) = 0$ for $r < \sigma$.
$g(r)$ has a discontinuity at $r = \sigma$ and approaches a value larger than unity in the limit $r \to \sigma + 0$, indicating the presence of effective attraction as a many-body effect.
For large $r / \sigma$ values, $g(r)$ is expected to approach unity, given that the hard-core repulsion is a short-range phenomenon. 
For moderate densities of $bn$, accurate expressions for $g(r)$ of the hard-sphere system can be obtained based on the solution of the Percus-Yevick equation~\cite{PhysRevLett.10.321,doi:10.1063/1.1734272,PhysRevA.43.5418}.
This can be used to validate the performance of the \fsampler.

\begin{figure}[t]
  \centering
  \includegraphics[width=.49\textwidth]{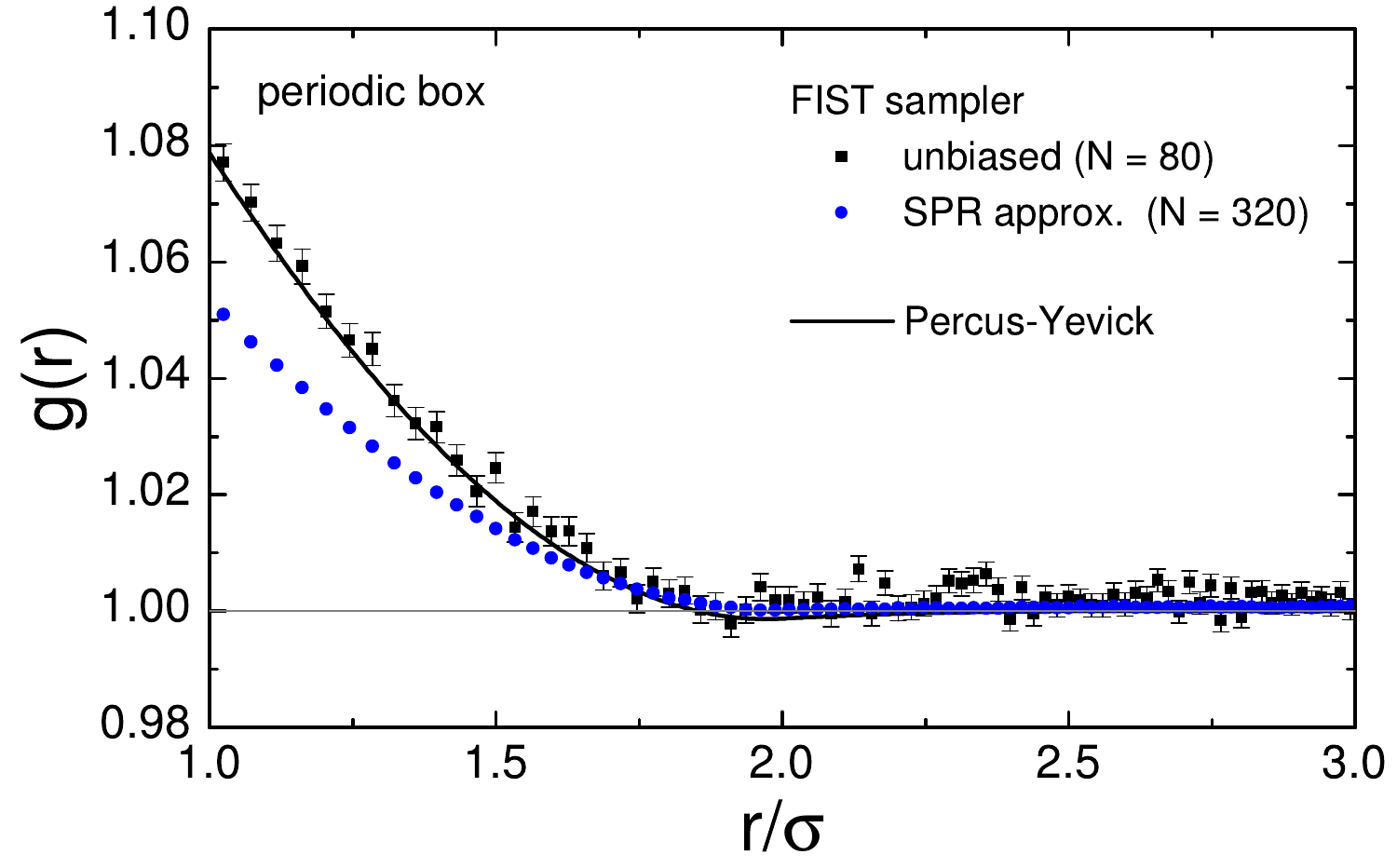}
  \caption{
  Radial distribution function $g(r)$ resulting from \fsampler~for a system of particles in a periodic box at density $bn = 0.12$ with~(blue circles) and without~(black squares) the application of the SPR approximation. The black line corresponds to $g(r)$ of a hard-sphere system calculated through the Percus-Yevick equation~\cite{PhysRevA.43.5418}.
  }
  \label{fig:HS.RDF}
\end{figure}

Figure~\ref{fig:HS.RDF} depicts $g(r)$ in a range $1 < r/\sigma < 3$ resulting from \fsampler~for a system of particles in a periodic box at density $bn = 0.12$, which was obtained by binning the relative distances of the sampled particles.
Blue circles correspond to the case where the SPR approximation was applied, whereas black squares show the unbiased calculation without approximations. In the former case $N = 320$ particles were sampled in each configuration, while in the latter case, only $N = 80$ particles were used due to slower computational performance.

The results are compared with the analytical expectation for a hard-sphere system based on the Percus-Yevick approximation, which is expected to be highly accurate for density as low as $bn = 0.12$~\cite{PhysRevA.43.5418}.
The full calculation is in quantitative agreement with the analytical expectations, indicating that \fsampler~correctly reproduces the spatial correlations between particles that interact through the hard-core potential~\eqref{eq:VHC}.
The SPR approximation underestimates $g(r)$ at $1 < r/\sigma \lesssim 1.5$, indicating that it does not capture the full strength of the many-body effective attraction effect.
At the same time, the SPR approximation reproduces the qualitative structure of $g(r)$, and, as discussed in Sec.~\ref{sec:SPR}, reproduces accurately the behavior of cumulants of particle number distribution for densities considered in this study.

\section{Comparison of \fsampler~with analytic calculations at $\sNN = 2.4$~GeV}
\label{app:SAMHADES}

\begin{figure}[t]
  \centering
  \includegraphics[width=.49\textwidth]{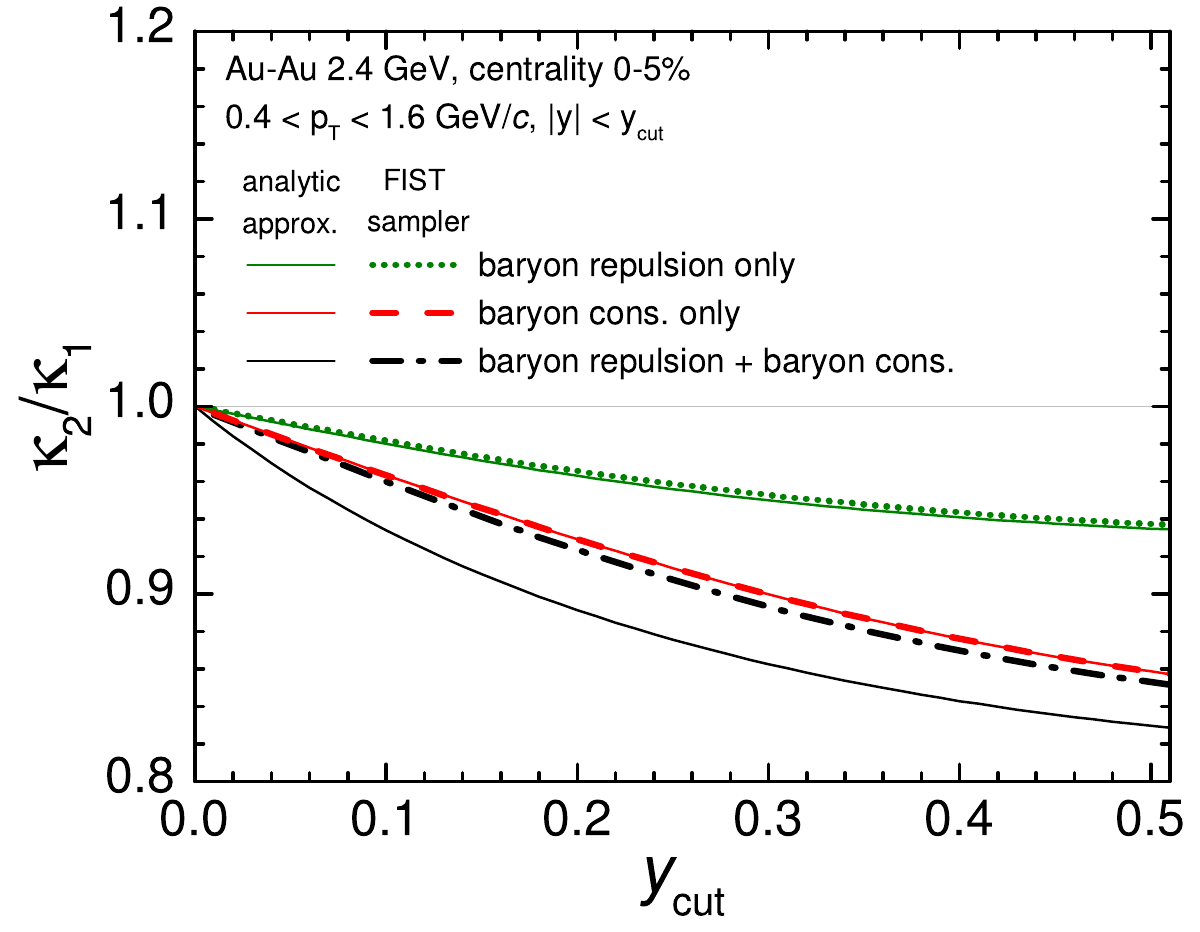}
  \caption{
   The scaled variance of the proton number distribution in the experimental acceptance of the HADES Collaboration as a function of the rapidity cut $y_{\rm cut}$. The green~(top) lines correspond to calculations incorporating the hard-core repulsion among baryons but not baryon conservation. The calculations shown by the red~(middle) lines implement exact baryon conservation but not baryon repulsion. The black~(bottom) lines show calculations with both the baryon repulsion and conservation included. The differently styled lines correspond to the results obtained using \fsampler~while the solid lines correspond to the analytic method from Ref.~\cite{Vovchenko:2022szk}.
  }
  \label{fig:HADES.SAMtest}
\end{figure}

Here a comparison of \fsampler~calculations with the analytic method recently developed in Ref.~\cite{Vovchenko:2022szk} is performed.
This allows one to verify the accuracy of the analytic approximations of Ref.~\cite{Vovchenko:2022szk}.
In particular, Ref.~\cite{Vovchenko:2022szk} used a method called SAM-2.0~\cite{Vovchenko:2021yen} to perform a correction for exact baryon conservation. As discussed in Ref.~\cite{Vovchenko:2021yen}, the method might overestimate the effect of dynamical correlations among particles, in this case, the effect of short-range hard-core repulsion, at low collision energies such as $\sNN = 2.4$~GeV at SIS-GSI.
By comparing to \fsampler, which is free of these issues, one can quantify the accuracy of SAM-2.0 in this collision energy regime.

The corresponding results are shown in Fig.~\ref{fig:HADES.SAMtest}.
One sees good agreement between Monte Carlo and analytic calculations for the cases when baryon repulsion is incorporated but not conservation~(green lines) or when baryon conservation is implemented but not repulsion~(red lines).
This validates the analytic approach of Ref.~\cite{Vovchenko:2022szk} in these regimes, in particular, SAM-2.0 used in~\cite{Vovchenko:2022szk} to correct for baryon conservation is accurate in the limit of the ideal gas, as expected.

When both the baryon repulsion and conservation are incorporated, visible differences between the \fsampler~and the analytic method occur. The analytic method, based on SAM-2.0, overestimates the suppression of $\kappa_2/\kappa_1$ compared to the Monte Carlo sampling. This is an artifact of the SAM-2.0 approximation, which assumes that there are no correlations between particles inside and outside the acceptance apart from the baryon conservation. As discussed in~\cite{Vovchenko:2021yen}, this leads to an overestimation of the correlations present in the grand-canonical limit, namely the effect of hard-core repulsion in this study. 
A similar observation applies to the higher-order cumulant ratios $\kappa_3/\kappa_2$ and $\kappa_4/\kappa_2$ that are not shown in Fig.~\ref{fig:HADES.SAMtest}.

As follows from the \fsampler~calculations in Fig.~\ref{fig:HADES.SAMtest}, the presence of baryon repulsion in addition to baryon conservation leads to only a minor further suppression of proton cumulant ratios, i.e., the effect of hard-core repulsion in the cumulants appears to be essentially washed out by baryon conservation.
This appears to be a consequence of the reduced collective flow effect at HADES energies compared to higher energies, which dilutes the space-momentum correlation, meaning that the excluded volume effect in the coordinate space is almost washed out in the momentum space where the measurements are performed. 
As shown in recent molecular dynamics simulations~\cite{Kuznietsov:2022pcn}, even critical fluctuations may be washed out when analyzed in momentum space if exact global conservation laws are enforced, and no collective expansion is imposed.

\bibliography{CFEV}

\begin{thebibliography}{70}%
\makeatletter
\providecommand \@ifxundefined [1]{%
 \@ifx{#1\undefined}
}%
\providecommand \@ifnum [1]{%
 \ifnum #1\expandafter \@firstoftwo
 \else \expandafter \@secondoftwo
 \fi
}%
\providecommand \@ifx [1]{%
 \ifx #1\expandafter \@firstoftwo
 \else \expandafter \@secondoftwo
 \fi
}%
\providecommand \natexlab [1]{#1}%
\providecommand \enquote  [1]{``#1''}%
\providecommand \bibnamefont  [1]{#1}%
\providecommand \bibfnamefont [1]{#1}%
\providecommand \citenamefont [1]{#1}%
\providecommand \href@noop [0]{\@secondoftwo}%
\providecommand \href [0]{\begingroup \@sanitize@url \@href}%
\providecommand \@href[1]{\@@startlink{#1}\@@href}%
\providecommand \@@href[1]{\endgroup#1\@@endlink}%
\providecommand \@sanitize@url [0]{\catcode `\\12\catcode `\$12\catcode
  `\&12\catcode `\#12\catcode `\^12\catcode `\_12\catcode `\%12\relax}%
\providecommand \@@startlink[1]{}%
\providecommand \@@endlink[0]{}%
\providecommand \url  [0]{\begingroup\@sanitize@url \@url }%
\providecommand \@url [1]{\endgroup\@href {#1}{\urlprefix }}%
\providecommand \urlprefix  [0]{URL }%
\providecommand \Eprint [0]{\href }%
\providecommand \doibase [0]{http://dx.doi.org/}%
\providecommand \selectlanguage [0]{\@gobble}%
\providecommand \bibinfo  [0]{\@secondoftwo}%
\providecommand \bibfield  [0]{\@secondoftwo}%
\providecommand \translation [1]{[#1]}%
\providecommand \BibitemOpen [0]{}%
\providecommand \bibitemStop [0]{}%
\providecommand \bibitemNoStop [0]{.\EOS\space}%
\providecommand \EOS [0]{\spacefactor3000\relax}%
\providecommand \BibitemShut  [1]{\csname bibitem#1\endcsname}%
\let\auto@bib@innerbib\@empty
%</preamble>
\bibitem [{\citenamefont {Bzdak}\ \emph {et~al.}(2020)\citenamefont {Bzdak},
  \citenamefont {Esumi}, \citenamefont {Koch}, \citenamefont {Liao},
  \citenamefont {Stephanov},\ and\ \citenamefont {Xu}}]{Bzdak:2019pkr}%
  \BibitemOpen
  \bibfield  {author} {\bibinfo {author} {\bibfnamefont {A.}~\bibnamefont
  {Bzdak}}, \bibinfo {author} {\bibfnamefont {S.}~\bibnamefont {Esumi}},
  \bibinfo {author} {\bibfnamefont {V.}~\bibnamefont {Koch}}, \bibinfo {author}
  {\bibfnamefont {J.}~\bibnamefont {Liao}}, \bibinfo {author} {\bibfnamefont
  {M.}~\bibnamefont {Stephanov}}, \ and\ \bibinfo {author} {\bibfnamefont
  {N.}~\bibnamefont {Xu}},\ }\href {\doibase 10.1016/j.physrep.2020.01.005}
  {\bibfield  {journal} {\bibinfo  {journal} {Phys. Rept.}\ }\textbf {\bibinfo
  {volume} {853}},\ \bibinfo {pages} {1} (\bibinfo {year} {2020})},\ \Eprint
  {http://arxiv.org/abs/1906.00936} {arXiv:1906.00936 [nucl-th]} \BibitemShut
  {NoStop}%
\bibitem [{\citenamefont {Shen}\ and\ \citenamefont
  {Yan}(2020)}]{Shen:2020mgh}%
  \BibitemOpen
  \bibfield  {author} {\bibinfo {author} {\bibfnamefont {C.}~\bibnamefont
  {Shen}}\ and\ \bibinfo {author} {\bibfnamefont {L.}~\bibnamefont {Yan}},\
  }\href {\doibase 10.1007/s41365-020-00829-z} {\bibfield  {journal} {\bibinfo
  {journal} {Nucl. Sci. Tech.}\ }\textbf {\bibinfo {volume} {31}},\ \bibinfo
  {pages} {122} (\bibinfo {year} {2020})},\ \Eprint
  {http://arxiv.org/abs/2010.12377} {arXiv:2010.12377 [nucl-th]} \BibitemShut
  {NoStop}%
\bibitem [{\citenamefont {Kisiel}\ \emph {et~al.}(2006)\citenamefont {Kisiel},
  \citenamefont {Taluc}, \citenamefont {Broniowski},\ and\ \citenamefont
  {Florkowski}}]{Kisiel:2005hn}%
  \BibitemOpen
  \bibfield  {author} {\bibinfo {author} {\bibfnamefont {A.}~\bibnamefont
  {Kisiel}}, \bibinfo {author} {\bibfnamefont {T.}~\bibnamefont {Taluc}},
  \bibinfo {author} {\bibfnamefont {W.}~\bibnamefont {Broniowski}}, \ and\
  \bibinfo {author} {\bibfnamefont {W.}~\bibnamefont {Florkowski}},\ }\href
  {\doibase 10.1016/j.cpc.2005.11.010} {\bibfield  {journal} {\bibinfo
  {journal} {Comput. Phys. Commun.}\ }\textbf {\bibinfo {volume} {174}},\
  \bibinfo {pages} {669} (\bibinfo {year} {2006})},\ \Eprint
  {http://arxiv.org/abs/nucl-th/0504047} {arXiv:nucl-th/0504047} \BibitemShut
  {NoStop}%
\bibitem [{\citenamefont {Pratt}\ and\ \citenamefont
  {Torrieri}(2010)}]{Pratt:2010jt}%
  \BibitemOpen
  \bibfield  {author} {\bibinfo {author} {\bibfnamefont {S.}~\bibnamefont
  {Pratt}}\ and\ \bibinfo {author} {\bibfnamefont {G.}~\bibnamefont
  {Torrieri}},\ }\href {\doibase 10.1103/PhysRevC.82.044901} {\bibfield
  {journal} {\bibinfo  {journal} {Phys. Rev. C}\ }\textbf {\bibinfo {volume}
  {82}},\ \bibinfo {pages} {044901} (\bibinfo {year} {2010})},\ \Eprint
  {http://arxiv.org/abs/1003.0413} {arXiv:1003.0413 [nucl-th]} \BibitemShut
  {NoStop}%
\bibitem [{\citenamefont {Shen}\ \emph {et~al.}(2016)\citenamefont {Shen},
  \citenamefont {Qiu}, \citenamefont {Song}, \citenamefont {Bernhard},
  \citenamefont {Bass},\ and\ \citenamefont {Heinz}}]{Shen:2014vra}%
  \BibitemOpen
  \bibfield  {author} {\bibinfo {author} {\bibfnamefont {C.}~\bibnamefont
  {Shen}}, \bibinfo {author} {\bibfnamefont {Z.}~\bibnamefont {Qiu}}, \bibinfo
  {author} {\bibfnamefont {H.}~\bibnamefont {Song}}, \bibinfo {author}
  {\bibfnamefont {J.}~\bibnamefont {Bernhard}}, \bibinfo {author}
  {\bibfnamefont {S.}~\bibnamefont {Bass}}, \ and\ \bibinfo {author}
  {\bibfnamefont {U.}~\bibnamefont {Heinz}},\ }\href {\doibase
  10.1016/j.cpc.2015.08.039} {\bibfield  {journal} {\bibinfo  {journal}
  {Comput. Phys. Commun.}\ }\textbf {\bibinfo {volume} {199}},\ \bibinfo
  {pages} {61} (\bibinfo {year} {2016})},\ \Eprint
  {http://arxiv.org/abs/1409.8164} {arXiv:1409.8164 [nucl-th]} \BibitemShut
  {NoStop}%
\bibitem [{\citenamefont {Karpenko}\ \emph {et~al.}(2015)\citenamefont
  {Karpenko}, \citenamefont {Huovinen}, \citenamefont {Petersen},\ and\
  \citenamefont {Bleicher}}]{Karpenko:2015xea}%
  \BibitemOpen
  \bibfield  {author} {\bibinfo {author} {\bibfnamefont {I.~A.}\ \bibnamefont
  {Karpenko}}, \bibinfo {author} {\bibfnamefont {P.}~\bibnamefont {Huovinen}},
  \bibinfo {author} {\bibfnamefont {H.}~\bibnamefont {Petersen}}, \ and\
  \bibinfo {author} {\bibfnamefont {M.}~\bibnamefont {Bleicher}},\ }\href
  {\doibase 10.1103/PhysRevC.91.064901} {\bibfield  {journal} {\bibinfo
  {journal} {Phys. Rev. C}\ }\textbf {\bibinfo {volume} {91}},\ \bibinfo
  {pages} {064901} (\bibinfo {year} {2015})},\ \Eprint
  {http://arxiv.org/abs/1502.01978} {arXiv:1502.01978 [nucl-th]} \BibitemShut
  {NoStop}%
\bibitem [{\citenamefont {Bernhard}(2018)}]{Bernhard:2018hnz}%
  \BibitemOpen
  \bibfield  {author} {\bibinfo {author} {\bibfnamefont {J.~E.}\ \bibnamefont
  {Bernhard}},\ }\emph {\bibinfo {title} {{Bayesian parameter estimation for
  relativistic heavy-ion collisions}}},\ \href@noop {} {Ph.D. thesis},\
  \bibinfo  {school} {Duke U.} (\bibinfo {year} {2018}),\ \Eprint
  {http://arxiv.org/abs/1804.06469} {arXiv:1804.06469 [nucl-th]} \BibitemShut
  {NoStop}%
\bibitem [{\citenamefont {Becattini}\ and\ \citenamefont
  {Ferroni}(2004{\natexlab{a}})}]{Becattini:2003ft}%
  \BibitemOpen
  \bibfield  {author} {\bibinfo {author} {\bibfnamefont {F.}~\bibnamefont
  {Becattini}}\ and\ \bibinfo {author} {\bibfnamefont {L.}~\bibnamefont
  {Ferroni}},\ }\href {\doibase 10.1140/epjc/s2004-01850-1} {\bibfield
  {journal} {\bibinfo  {journal} {Eur. Phys. J. C}\ }\textbf {\bibinfo {volume}
  {35}},\ \bibinfo {pages} {243} (\bibinfo {year} {2004}{\natexlab{a}})},\
  \Eprint {http://arxiv.org/abs/hep-ph/0307061} {arXiv:hep-ph/0307061}
  \BibitemShut {NoStop}%
\bibitem [{\citenamefont {Becattini}\ and\ \citenamefont
  {Ferroni}(2004{\natexlab{b}})}]{Becattini:2004rq}%
  \BibitemOpen
  \bibfield  {author} {\bibinfo {author} {\bibfnamefont {F.}~\bibnamefont
  {Becattini}}\ and\ \bibinfo {author} {\bibfnamefont {L.}~\bibnamefont
  {Ferroni}},\ }\href {\doibase 10.1140/epjc/s10052-010-1243-4} {\bibfield
  {journal} {\bibinfo  {journal} {Eur. Phys. J. C}\ }\textbf {\bibinfo {volume}
  {38}},\ \bibinfo {pages} {225} (\bibinfo {year} {2004}{\natexlab{b}})},\
  \bibinfo {note} {[Erratum: Eur.Phys.J. 66, 341 (2010)]},\ \Eprint
  {http://arxiv.org/abs/hep-ph/0407117} {arXiv:hep-ph/0407117} \BibitemShut
  {NoStop}%
\bibitem [{\citenamefont {Schwarz}\ \emph {et~al.}(2018)\citenamefont
  {Schwarz}, \citenamefont {Oliinychenko}, \citenamefont {Pang}, \citenamefont
  {Ryu},\ and\ \citenamefont {Petersen}}]{Schwarz:2017bdg}%
  \BibitemOpen
  \bibfield  {author} {\bibinfo {author} {\bibfnamefont {C.}~\bibnamefont
  {Schwarz}}, \bibinfo {author} {\bibfnamefont {D.}~\bibnamefont
  {Oliinychenko}}, \bibinfo {author} {\bibfnamefont {L.~G.}\ \bibnamefont
  {Pang}}, \bibinfo {author} {\bibfnamefont {S.}~\bibnamefont {Ryu}}, \ and\
  \bibinfo {author} {\bibfnamefont {H.}~\bibnamefont {Petersen}},\ }\href
  {\doibase 10.1088/1361-6471/aa90eb} {\bibfield  {journal} {\bibinfo
  {journal} {J. Phys. G}\ }\textbf {\bibinfo {volume} {45}},\ \bibinfo {pages}
  {015001} (\bibinfo {year} {2018})},\ \Eprint
  {http://arxiv.org/abs/1707.07026} {arXiv:1707.07026 [hep-ph]} \BibitemShut
  {NoStop}%
\bibitem [{\citenamefont {Oliinychenko}\ and\ \citenamefont
  {Koch}(2019)}]{Oliinychenko:2019zfk}%
  \BibitemOpen
  \bibfield  {author} {\bibinfo {author} {\bibfnamefont {D.}~\bibnamefont
  {Oliinychenko}}\ and\ \bibinfo {author} {\bibfnamefont {V.}~\bibnamefont
  {Koch}},\ }\href {\doibase 10.1103/PhysRevLett.123.182302} {\bibfield
  {journal} {\bibinfo  {journal} {Phys. Rev. Lett.}\ }\textbf {\bibinfo
  {volume} {123}},\ \bibinfo {pages} {182302} (\bibinfo {year} {2019})},\
  \Eprint {http://arxiv.org/abs/1902.09775} {arXiv:1902.09775 [hep-ph]}
  \BibitemShut {NoStop}%
\bibitem [{\citenamefont {Oliinychenko}\ \emph {et~al.}(2020)\citenamefont
  {Oliinychenko}, \citenamefont {Shi},\ and\ \citenamefont
  {Koch}}]{Oliinychenko:2020cmr}%
  \BibitemOpen
  \bibfield  {author} {\bibinfo {author} {\bibfnamefont {D.}~\bibnamefont
  {Oliinychenko}}, \bibinfo {author} {\bibfnamefont {S.}~\bibnamefont {Shi}}, \
  and\ \bibinfo {author} {\bibfnamefont {V.}~\bibnamefont {Koch}},\ }\href
  {\doibase 10.1103/PhysRevC.102.034904} {\bibfield  {journal} {\bibinfo
  {journal} {Phys. Rev. C}\ }\textbf {\bibinfo {volume} {102}},\ \bibinfo
  {pages} {034904} (\bibinfo {year} {2020})},\ \Eprint
  {http://arxiv.org/abs/2001.08176} {arXiv:2001.08176 [hep-ph]} \BibitemShut
  {NoStop}%
\bibitem [{\citenamefont {Begun}\ \emph {et~al.}(2004)\citenamefont {Begun},
  \citenamefont {Gazdzicki}, \citenamefont {Gorenstein},\ and\ \citenamefont
  {Zozulya}}]{Begun:2004gs}%
  \BibitemOpen
  \bibfield  {author} {\bibinfo {author} {\bibfnamefont {V.~V.}\ \bibnamefont
  {Begun}}, \bibinfo {author} {\bibfnamefont {M.}~\bibnamefont {Gazdzicki}},
  \bibinfo {author} {\bibfnamefont {M.~I.}\ \bibnamefont {Gorenstein}}, \ and\
  \bibinfo {author} {\bibfnamefont {O.~S.}\ \bibnamefont {Zozulya}},\ }\href
  {\doibase 10.1103/PhysRevC.70.034901} {\bibfield  {journal} {\bibinfo
  {journal} {Phys. Rev. C}\ }\textbf {\bibinfo {volume} {70}},\ \bibinfo
  {pages} {034901} (\bibinfo {year} {2004})},\ \Eprint
  {http://arxiv.org/abs/nucl-th/0404056} {arXiv:nucl-th/0404056} \BibitemShut
  {NoStop}%
\bibitem [{\citenamefont {Begun}\ \emph {et~al.}(2006)\citenamefont {Begun},
  \citenamefont {Gorenstein}, \citenamefont {Hauer}, \citenamefont
  {Konchakovski},\ and\ \citenamefont {Zozulya}}]{Begun:2006jf}%
  \BibitemOpen
  \bibfield  {author} {\bibinfo {author} {\bibfnamefont {V.~V.}\ \bibnamefont
  {Begun}}, \bibinfo {author} {\bibfnamefont {M.~I.}\ \bibnamefont
  {Gorenstein}}, \bibinfo {author} {\bibfnamefont {M.}~\bibnamefont {Hauer}},
  \bibinfo {author} {\bibfnamefont {V.~P.}\ \bibnamefont {Konchakovski}}, \
  and\ \bibinfo {author} {\bibfnamefont {O.~S.}\ \bibnamefont {Zozulya}},\
  }\href {\doibase 10.1103/PhysRevC.74.044903} {\bibfield  {journal} {\bibinfo
  {journal} {Phys. Rev. C}\ }\textbf {\bibinfo {volume} {74}},\ \bibinfo
  {pages} {044903} (\bibinfo {year} {2006})},\ \Eprint
  {http://arxiv.org/abs/nucl-th/0606036} {arXiv:nucl-th/0606036} \BibitemShut
  {NoStop}%
\bibitem [{\citenamefont {Braun-Munzinger}\ \emph {et~al.}(2021)\citenamefont
  {Braun-Munzinger}, \citenamefont {Friman}, \citenamefont {Redlich},
  \citenamefont {Rustamov},\ and\ \citenamefont
  {Stachel}}]{Braun-Munzinger:2020jbk}%
  \BibitemOpen
  \bibfield  {author} {\bibinfo {author} {\bibfnamefont {P.}~\bibnamefont
  {Braun-Munzinger}}, \bibinfo {author} {\bibfnamefont {B.}~\bibnamefont
  {Friman}}, \bibinfo {author} {\bibfnamefont {K.}~\bibnamefont {Redlich}},
  \bibinfo {author} {\bibfnamefont {A.}~\bibnamefont {Rustamov}}, \ and\
  \bibinfo {author} {\bibfnamefont {J.}~\bibnamefont {Stachel}},\ }\href
  {\doibase 10.1016/j.nuclphysa.2021.122141} {\bibfield  {journal} {\bibinfo
  {journal} {Nucl. Phys. A}\ }\textbf {\bibinfo {volume} {1008}},\ \bibinfo
  {pages} {122141} (\bibinfo {year} {2021})},\ \Eprint
  {http://arxiv.org/abs/2007.02463} {arXiv:2007.02463 [nucl-th]} \BibitemShut
  {NoStop}%
\bibitem [{\citenamefont {Vovchenko}\ and\ \citenamefont
  {Koch}(2021)}]{Vovchenko:2020kwg}%
  \BibitemOpen
  \bibfield  {author} {\bibinfo {author} {\bibfnamefont {V.}~\bibnamefont
  {Vovchenko}}\ and\ \bibinfo {author} {\bibfnamefont {V.}~\bibnamefont
  {Koch}},\ }\href {\doibase 10.1103/PhysRevC.103.044903} {\bibfield  {journal}
  {\bibinfo  {journal} {Phys. Rev. C}\ }\textbf {\bibinfo {volume} {103}},\
  \bibinfo {pages} {044903} (\bibinfo {year} {2021})},\ \Eprint
  {http://arxiv.org/abs/2012.09954} {arXiv:2012.09954 [hep-ph]} \BibitemShut
  {NoStop}%
\bibitem [{\citenamefont {Vovchenko}\ \emph {et~al.}(2022)\citenamefont
  {Vovchenko}, \citenamefont {Koch},\ and\ \citenamefont
  {Shen}}]{Vovchenko:2021kxx}%
  \BibitemOpen
  \bibfield  {author} {\bibinfo {author} {\bibfnamefont {V.}~\bibnamefont
  {Vovchenko}}, \bibinfo {author} {\bibfnamefont {V.}~\bibnamefont {Koch}}, \
  and\ \bibinfo {author} {\bibfnamefont {C.}~\bibnamefont {Shen}},\ }\href
  {\doibase 10.1103/PhysRevC.105.014904} {\bibfield  {journal} {\bibinfo
  {journal} {Phys. Rev. C}\ }\textbf {\bibinfo {volume} {105}},\ \bibinfo
  {pages} {014904} (\bibinfo {year} {2022})},\ \Eprint
  {http://arxiv.org/abs/2107.00163} {arXiv:2107.00163 [hep-ph]} \BibitemShut
  {NoStop}%
\bibitem [{\citenamefont {Yen}\ \emph {et~al.}(1997)\citenamefont {Yen},
  \citenamefont {Gorenstein}, \citenamefont {Greiner},\ and\ \citenamefont
  {Yang}}]{Yen:1997rv}%
  \BibitemOpen
  \bibfield  {author} {\bibinfo {author} {\bibfnamefont {G.~D.}\ \bibnamefont
  {Yen}}, \bibinfo {author} {\bibfnamefont {M.~I.}\ \bibnamefont {Gorenstein}},
  \bibinfo {author} {\bibfnamefont {W.}~\bibnamefont {Greiner}}, \ and\
  \bibinfo {author} {\bibfnamefont {S.-N.}\ \bibnamefont {Yang}},\ }\href
  {\doibase 10.1103/PhysRevC.56.2210} {\bibfield  {journal} {\bibinfo
  {journal} {Phys. Rev. C}\ }\textbf {\bibinfo {volume} {56}},\ \bibinfo
  {pages} {2210} (\bibinfo {year} {1997})},\ \Eprint
  {http://arxiv.org/abs/nucl-th/9711062} {arXiv:nucl-th/9711062} \BibitemShut
  {NoStop}%
\bibitem [{\citenamefont {Satarov}\ \emph {et~al.}(2017)\citenamefont
  {Satarov}, \citenamefont {Vovchenko}, \citenamefont {Alba}, \citenamefont
  {Gorenstein},\ and\ \citenamefont {Stoecker}}]{Satarov:2016peb}%
  \BibitemOpen
  \bibfield  {author} {\bibinfo {author} {\bibfnamefont {L.~M.}\ \bibnamefont
  {Satarov}}, \bibinfo {author} {\bibfnamefont {V.}~\bibnamefont {Vovchenko}},
  \bibinfo {author} {\bibfnamefont {P.}~\bibnamefont {Alba}}, \bibinfo {author}
  {\bibfnamefont {M.~I.}\ \bibnamefont {Gorenstein}}, \ and\ \bibinfo {author}
  {\bibfnamefont {H.}~\bibnamefont {Stoecker}},\ }\href {\doibase
  10.1103/PhysRevC.95.024902} {\bibfield  {journal} {\bibinfo  {journal} {Phys.
  Rev. C}\ }\textbf {\bibinfo {volume} {95}},\ \bibinfo {pages} {024902}
  (\bibinfo {year} {2017})},\ \Eprint {http://arxiv.org/abs/1610.08753}
  {arXiv:1610.08753 [nucl-th]} \BibitemShut {NoStop}%
\bibitem [{\citenamefont {Vovchenko}(2020)}]{Vovchenko:2020lju}%
  \BibitemOpen
  \bibfield  {author} {\bibinfo {author} {\bibfnamefont {V.}~\bibnamefont
  {Vovchenko}},\ }\href {\doibase 10.1142/S0218301320400029} {\bibfield
  {journal} {\bibinfo  {journal} {Int. J. Mod. Phys. E}\ }\textbf {\bibinfo
  {volume} {29}},\ \bibinfo {pages} {2040002} (\bibinfo {year} {2020})},\
  \Eprint {http://arxiv.org/abs/2004.06331} {arXiv:2004.06331 [nucl-th]}
  \BibitemShut {NoStop}%
\bibitem [{\citenamefont {Vovchenko}\ \emph
  {et~al.}(2017{\natexlab{a}})\citenamefont {Vovchenko}, \citenamefont
  {Gorenstein},\ and\ \citenamefont {Stoecker}}]{Vovchenko:2016rkn}%
  \BibitemOpen
  \bibfield  {author} {\bibinfo {author} {\bibfnamefont {V.}~\bibnamefont
  {Vovchenko}}, \bibinfo {author} {\bibfnamefont {M.~I.}\ \bibnamefont
  {Gorenstein}}, \ and\ \bibinfo {author} {\bibfnamefont {H.}~\bibnamefont
  {Stoecker}},\ }\href {\doibase 10.1103/PhysRevLett.118.182301} {\bibfield
  {journal} {\bibinfo  {journal} {Phys. Rev. Lett.}\ }\textbf {\bibinfo
  {volume} {118}},\ \bibinfo {pages} {182301} (\bibinfo {year}
  {2017}{\natexlab{a}})},\ \Eprint {http://arxiv.org/abs/1609.03975}
  {arXiv:1609.03975 [hep-ph]} \BibitemShut {NoStop}%
\bibitem [{\citenamefont {Vovchenko}\ \emph
  {et~al.}(2017{\natexlab{b}})\citenamefont {Vovchenko}, \citenamefont
  {Pasztor}, \citenamefont {Fodor}, \citenamefont {Katz},\ and\ \citenamefont
  {Stoecker}}]{Vovchenko:2017xad}%
  \BibitemOpen
  \bibfield  {author} {\bibinfo {author} {\bibfnamefont {V.}~\bibnamefont
  {Vovchenko}}, \bibinfo {author} {\bibfnamefont {A.}~\bibnamefont {Pasztor}},
  \bibinfo {author} {\bibfnamefont {Z.}~\bibnamefont {Fodor}}, \bibinfo
  {author} {\bibfnamefont {S.~D.}\ \bibnamefont {Katz}}, \ and\ \bibinfo
  {author} {\bibfnamefont {H.}~\bibnamefont {Stoecker}},\ }\href {\doibase
  10.1016/j.physletb.2017.10.042} {\bibfield  {journal} {\bibinfo  {journal}
  {Phys. Lett. B}\ }\textbf {\bibinfo {volume} {775}},\ \bibinfo {pages} {71}
  (\bibinfo {year} {2017}{\natexlab{b}})},\ \Eprint
  {http://arxiv.org/abs/1708.02852} {arXiv:1708.02852 [hep-ph]} \BibitemShut
  {NoStop}%
\bibitem [{\citenamefont {Huovinen}\ and\ \citenamefont
  {Petreczky}(2018)}]{Huovinen:2017ogf}%
  \BibitemOpen
  \bibfield  {author} {\bibinfo {author} {\bibfnamefont {P.}~\bibnamefont
  {Huovinen}}\ and\ \bibinfo {author} {\bibfnamefont {P.}~\bibnamefont
  {Petreczky}},\ }\href {\doibase 10.1016/j.physletb.2017.12.001} {\bibfield
  {journal} {\bibinfo  {journal} {Phys. Lett. B}\ }\textbf {\bibinfo {volume}
  {777}},\ \bibinfo {pages} {125} (\bibinfo {year} {2018})},\ \Eprint
  {http://arxiv.org/abs/1708.00879} {arXiv:1708.00879 [hep-ph]} \BibitemShut
  {NoStop}%
\bibitem [{\citenamefont {Karthein}\ \emph {et~al.}(2021)\citenamefont
  {Karthein}, \citenamefont {Koch}, \citenamefont {Ratti},\ and\ \citenamefont
  {Vovchenko}}]{Karthein:2021cmb}%
  \BibitemOpen
  \bibfield  {author} {\bibinfo {author} {\bibfnamefont {J.~M.}\ \bibnamefont
  {Karthein}}, \bibinfo {author} {\bibfnamefont {V.}~\bibnamefont {Koch}},
  \bibinfo {author} {\bibfnamefont {C.}~\bibnamefont {Ratti}}, \ and\ \bibinfo
  {author} {\bibfnamefont {V.}~\bibnamefont {Vovchenko}},\ }\href {\doibase
  10.1103/PhysRevD.104.094009} {\bibfield  {journal} {\bibinfo  {journal}
  {Phys. Rev. D}\ }\textbf {\bibinfo {volume} {104}},\ \bibinfo {pages}
  {094009} (\bibinfo {year} {2021})},\ \Eprint
  {http://arxiv.org/abs/2107.00588} {arXiv:2107.00588 [nucl-th]} \BibitemShut
  {NoStop}%
\bibitem [{\citenamefont {Bollweg}\ \emph {et~al.}(2021)\citenamefont
  {Bollweg}, \citenamefont {Goswami}, \citenamefont {Kaczmarek}, \citenamefont
  {Karsch}, \citenamefont {Mukherjee}, \citenamefont {Petreczky}, \citenamefont
  {Schmidt},\ and\ \citenamefont {Scior}}]{Bollweg:2021vqf}%
  \BibitemOpen
  \bibfield  {author} {\bibinfo {author} {\bibfnamefont {D.}~\bibnamefont
  {Bollweg}}, \bibinfo {author} {\bibfnamefont {J.}~\bibnamefont {Goswami}},
  \bibinfo {author} {\bibfnamefont {O.}~\bibnamefont {Kaczmarek}}, \bibinfo
  {author} {\bibfnamefont {F.}~\bibnamefont {Karsch}}, \bibinfo {author}
  {\bibfnamefont {S.}~\bibnamefont {Mukherjee}}, \bibinfo {author}
  {\bibfnamefont {P.}~\bibnamefont {Petreczky}}, \bibinfo {author}
  {\bibfnamefont {C.}~\bibnamefont {Schmidt}}, \ and\ \bibinfo {author}
  {\bibfnamefont {P.}~\bibnamefont {Scior}} (\bibinfo {collaboration}
  {HotQCD}),\ }\href {\doibase 10.1103/PhysRevD.104.074512} {\bibfield
  {journal} {\bibinfo  {journal} {Phys. Rev. D}\ }\textbf {\bibinfo {volume}
  {104}} (\bibinfo {year} {2021}),\ 10.1103/PhysRevD.104.074512},\ \Eprint
  {http://arxiv.org/abs/2107.10011} {arXiv:2107.10011 [hep-lat]} \BibitemShut
  {NoStop}%
\bibitem [{\citenamefont {Stephanov}(2009)}]{Stephanov:2008qz}%
  \BibitemOpen
  \bibfield  {author} {\bibinfo {author} {\bibfnamefont {M.~A.}\ \bibnamefont
  {Stephanov}},\ }\href {\doibase 10.1103/PhysRevLett.102.032301} {\bibfield
  {journal} {\bibinfo  {journal} {Phys. Rev. Lett.}\ }\textbf {\bibinfo
  {volume} {102}},\ \bibinfo {pages} {032301} (\bibinfo {year} {2009})},\
  \Eprint {http://arxiv.org/abs/0809.3450} {arXiv:0809.3450 [hep-ph]}
  \BibitemShut {NoStop}%
\bibitem [{\citenamefont {Ling}\ and\ \citenamefont
  {Stephanov}(2016)}]{Ling:2015yau}%
  \BibitemOpen
  \bibfield  {author} {\bibinfo {author} {\bibfnamefont {B.}~\bibnamefont
  {Ling}}\ and\ \bibinfo {author} {\bibfnamefont {M.~A.}\ \bibnamefont
  {Stephanov}},\ }\href {\doibase 10.1103/PhysRevC.93.034915} {\bibfield
  {journal} {\bibinfo  {journal} {Phys. Rev. C}\ }\textbf {\bibinfo {volume}
  {93}},\ \bibinfo {pages} {034915} (\bibinfo {year} {2016})},\ \Eprint
  {http://arxiv.org/abs/1512.09125} {arXiv:1512.09125 [nucl-th]} \BibitemShut
  {NoStop}%
\bibitem [{\citenamefont {Pradeep}\ \emph {et~al.}(2022)\citenamefont
  {Pradeep}, \citenamefont {Rajagopal}, \citenamefont {Stephanov},\ and\
  \citenamefont {Yin}}]{Pradeep:2022mkf}%
  \BibitemOpen
  \bibfield  {author} {\bibinfo {author} {\bibfnamefont {M.}~\bibnamefont
  {Pradeep}}, \bibinfo {author} {\bibfnamefont {K.}~\bibnamefont {Rajagopal}},
  \bibinfo {author} {\bibfnamefont {M.}~\bibnamefont {Stephanov}}, \ and\
  \bibinfo {author} {\bibfnamefont {Y.}~\bibnamefont {Yin}},\ }\href {\doibase
  10.1103/PhysRevD.106.036017} {\bibfield  {journal} {\bibinfo  {journal}
  {Phys. Rev. D}\ }\textbf {\bibinfo {volume} {106}},\ \bibinfo {pages}
  {036017} (\bibinfo {year} {2022})},\ \Eprint
  {http://arxiv.org/abs/2204.00639} {arXiv:2204.00639 [hep-ph]} \BibitemShut
  {NoStop}%
\bibitem [{\citenamefont {Vovchenko}\ \emph
  {et~al.}(2020{\natexlab{a}})\citenamefont {Vovchenko}, \citenamefont
  {Savchuk}, \citenamefont {Poberezhnyuk}, \citenamefont {Gorenstein},\ and\
  \citenamefont {Koch}}]{Vovchenko:2020tsr}%
  \BibitemOpen
  \bibfield  {author} {\bibinfo {author} {\bibfnamefont {V.}~\bibnamefont
  {Vovchenko}}, \bibinfo {author} {\bibfnamefont {O.}~\bibnamefont {Savchuk}},
  \bibinfo {author} {\bibfnamefont {R.~V.}\ \bibnamefont {Poberezhnyuk}},
  \bibinfo {author} {\bibfnamefont {M.~I.}\ \bibnamefont {Gorenstein}}, \ and\
  \bibinfo {author} {\bibfnamefont {V.}~\bibnamefont {Koch}},\ }\href {\doibase
  10.1016/j.physletb.2020.135868} {\bibfield  {journal} {\bibinfo  {journal}
  {Phys. Lett. B}\ }\textbf {\bibinfo {volume} {811}},\ \bibinfo {pages}
  {135868} (\bibinfo {year} {2020}{\natexlab{a}})},\ \Eprint
  {http://arxiv.org/abs/2003.13905} {arXiv:2003.13905 [hep-ph]} \BibitemShut
  {NoStop}%
\bibitem [{\citenamefont {Gorenstein}\ \emph {et~al.}(2007)\citenamefont
  {Gorenstein}, \citenamefont {Hauer},\ and\ \citenamefont
  {Nikolajenko}}]{Gorenstein:2007ep}%
  \BibitemOpen
  \bibfield  {author} {\bibinfo {author} {\bibfnamefont {M.~I.}\ \bibnamefont
  {Gorenstein}}, \bibinfo {author} {\bibfnamefont {M.}~\bibnamefont {Hauer}}, \
  and\ \bibinfo {author} {\bibfnamefont {D.~O.}\ \bibnamefont {Nikolajenko}},\
  }\href {\doibase 10.1103/PhysRevC.76.024901} {\bibfield  {journal} {\bibinfo
  {journal} {Phys. Rev. C}\ }\textbf {\bibinfo {volume} {76}},\ \bibinfo
  {pages} {024901} (\bibinfo {year} {2007})},\ \Eprint
  {http://arxiv.org/abs/nucl-th/0702081} {arXiv:nucl-th/0702081} \BibitemShut
  {NoStop}%
\bibitem [{\citenamefont {Cooper}\ and\ \citenamefont
  {Frye}(1974)}]{Cooper:1974mv}%
  \BibitemOpen
  \bibfield  {author} {\bibinfo {author} {\bibfnamefont {F.}~\bibnamefont
  {Cooper}}\ and\ \bibinfo {author} {\bibfnamefont {G.}~\bibnamefont {Frye}},\
  }\href {\doibase 10.1103/PhysRevD.10.186} {\bibfield  {journal} {\bibinfo
  {journal} {Phys. Rev. D}\ }\textbf {\bibinfo {volume} {10}},\ \bibinfo
  {pages} {186} (\bibinfo {year} {1974})}\BibitemShut {NoStop}%
\bibitem [{\citenamefont {Vovchenko}\ \emph {et~al.}(2018)\citenamefont
  {Vovchenko}, \citenamefont {Gorenstein},\ and\ \citenamefont
  {Stoecker}}]{Vovchenko:2018cnf}%
  \BibitemOpen
  \bibfield  {author} {\bibinfo {author} {\bibfnamefont {V.}~\bibnamefont
  {Vovchenko}}, \bibinfo {author} {\bibfnamefont {M.~I.}\ \bibnamefont
  {Gorenstein}}, \ and\ \bibinfo {author} {\bibfnamefont {H.}~\bibnamefont
  {Stoecker}},\ }\href {\doibase 10.1103/PhysRevC.98.064909} {\bibfield
  {journal} {\bibinfo  {journal} {Phys. Rev. C}\ }\textbf {\bibinfo {volume}
  {98}},\ \bibinfo {pages} {064909} (\bibinfo {year} {2018})},\ \Eprint
  {http://arxiv.org/abs/1805.01402} {arXiv:1805.01402 [nucl-th]} \BibitemShut
  {NoStop}%
\bibitem [{\citenamefont {Shen}(2020)}]{MUSICinput}%
  \BibitemOpen
  \bibfield  {author} {\bibinfo {author} {\bibfnamefont {C.}~\bibnamefont
  {Shen}},\ }\href@noop {} {} (\bibinfo {year} {2020}),\ \bibinfo {note}
  {\href{https://drive.google.com/drive/folders/1DMml4IXXcilEZaaTpGF2HM\_2ICmeydpz?usp=sharing}{Link}
  [Online; accessed 09-May-2021]}\BibitemShut {NoStop}%
\bibitem [{\citenamefont {Gorenstein}\ \emph {et~al.}(1999)\citenamefont
  {Gorenstein}, \citenamefont {Kostyuk},\ and\ \citenamefont
  {Krivenko}}]{Gorenstein:1999ce}%
  \BibitemOpen
  \bibfield  {author} {\bibinfo {author} {\bibfnamefont {M.~I.}\ \bibnamefont
  {Gorenstein}}, \bibinfo {author} {\bibfnamefont {A.~P.}\ \bibnamefont
  {Kostyuk}}, \ and\ \bibinfo {author} {\bibfnamefont {Y.~D.}\ \bibnamefont
  {Krivenko}},\ }\href {\doibase 10.1088/0954-3899/25/9/102} {\bibfield
  {journal} {\bibinfo  {journal} {J. Phys. G}\ }\textbf {\bibinfo {volume}
  {25}},\ \bibinfo {pages} {L75} (\bibinfo {year} {1999})},\ \Eprint
  {http://arxiv.org/abs/nucl-th/9906068} {arXiv:nucl-th/9906068} \BibitemShut
  {NoStop}%
\bibitem [{\citenamefont {Vovchenko}\ and\ \citenamefont
  {Stoecker}(2019)}]{Vovchenko:2019pjl}%
  \BibitemOpen
  \bibfield  {author} {\bibinfo {author} {\bibfnamefont {V.}~\bibnamefont
  {Vovchenko}}\ and\ \bibinfo {author} {\bibfnamefont {H.}~\bibnamefont
  {Stoecker}},\ }\href {\doibase 10.1016/j.cpc.2019.06.024} {\bibfield
  {journal} {\bibinfo  {journal} {Comput. Phys. Commun.}\ }\textbf {\bibinfo
  {volume} {244}},\ \bibinfo {pages} {295} (\bibinfo {year} {2019})},\ \Eprint
  {http://arxiv.org/abs/1901.05249} {arXiv:1901.05249 [nucl-th]} \BibitemShut
  {NoStop}%
\bibitem [{\citenamefont {Vovchenko}\ \emph
  {et~al.}(2020{\natexlab{b}})\citenamefont {Vovchenko}, \citenamefont
  {Poberezhnyuk},\ and\ \citenamefont {Koch}}]{Vovchenko:2020gne}%
  \BibitemOpen
  \bibfield  {author} {\bibinfo {author} {\bibfnamefont {V.}~\bibnamefont
  {Vovchenko}}, \bibinfo {author} {\bibfnamefont {R.~V.}\ \bibnamefont
  {Poberezhnyuk}}, \ and\ \bibinfo {author} {\bibfnamefont {V.}~\bibnamefont
  {Koch}},\ }\href {\doibase 10.1007/JHEP10(2020)089} {\bibfield  {journal}
  {\bibinfo  {journal} {JHEP}\ }\textbf {\bibinfo {volume} {10}},\ \bibinfo
  {pages} {089} (\bibinfo {year} {2020}{\natexlab{b}})},\ \Eprint
  {http://arxiv.org/abs/2007.03850} {arXiv:2007.03850 [hep-ph]} \BibitemShut
  {NoStop}%
\bibitem [{\citenamefont {Acharya}\ \emph
  {et~al.}(2020{\natexlab{a}})\citenamefont {Acharya} \emph
  {et~al.}}]{Acharya:2019izy}%
  \BibitemOpen
  \bibfield  {author} {\bibinfo {author} {\bibfnamefont {S.}~\bibnamefont
  {Acharya}} \emph {et~al.} (\bibinfo {collaboration} {ALICE}),\ }\href
  {\doibase 10.1016/j.physletb.2020.135564} {\bibfield  {journal} {\bibinfo
  {journal} {Phys. Lett. B}\ }\textbf {\bibinfo {volume} {807}},\ \bibinfo
  {pages} {135564} (\bibinfo {year} {2020}{\natexlab{a}})},\ \Eprint
  {http://arxiv.org/abs/1910.14396} {arXiv:1910.14396 [nucl-ex]} \BibitemShut
  {NoStop}%
\bibitem [{\citenamefont {Mazeliauskas}\ and\ \citenamefont
  {Vislavicius}(2020)}]{Mazeliauskas:2019ifr}%
  \BibitemOpen
  \bibfield  {author} {\bibinfo {author} {\bibfnamefont {A.}~\bibnamefont
  {Mazeliauskas}}\ and\ \bibinfo {author} {\bibfnamefont {V.}~\bibnamefont
  {Vislavicius}},\ }\href {\doibase 10.1103/PhysRevC.101.014910} {\bibfield
  {journal} {\bibinfo  {journal} {Phys. Rev. C}\ }\textbf {\bibinfo {volume}
  {101}},\ \bibinfo {pages} {014910} (\bibinfo {year} {2020})},\ \Eprint
  {http://arxiv.org/abs/1907.11059} {arXiv:1907.11059 [hep-ph]} \BibitemShut
  {NoStop}%
\bibitem [{\citenamefont {Acharya}\ \emph
  {et~al.}(2020{\natexlab{b}})\citenamefont {Acharya} \emph
  {et~al.}}]{ALICE:2019nbs}%
  \BibitemOpen
  \bibfield  {author} {\bibinfo {author} {\bibfnamefont {S.}~\bibnamefont
  {Acharya}} \emph {et~al.} (\bibinfo {collaboration} {ALICE}),\ }\href
  {\doibase 10.1016/j.physletb.2020.135564} {\bibfield  {journal} {\bibinfo
  {journal} {Phys. Lett. B}\ }\textbf {\bibinfo {volume} {807}},\ \bibinfo
  {pages} {135564} (\bibinfo {year} {2020}{\natexlab{b}})},\ \Eprint
  {http://arxiv.org/abs/1910.14396} {arXiv:1910.14396 [nucl-ex]} \BibitemShut
  {NoStop}%
\bibitem [{\citenamefont {Abdallah}\ \emph {et~al.}(2021)\citenamefont
  {Abdallah} \emph {et~al.}}]{STAR:2021iop}%
  \BibitemOpen
  \bibfield  {author} {\bibinfo {author} {\bibfnamefont {M.}~\bibnamefont
  {Abdallah}} \emph {et~al.} (\bibinfo {collaboration} {STAR}),\ }\href
  {\doibase 10.1103/PhysRevC.104.024902} {\bibfield  {journal} {\bibinfo
  {journal} {Phys. Rev. C}\ }\textbf {\bibinfo {volume} {104}},\ \bibinfo
  {pages} {024902} (\bibinfo {year} {2021})},\ \Eprint
  {http://arxiv.org/abs/2101.12413} {arXiv:2101.12413 [nucl-ex]} \BibitemShut
  {NoStop}%
\bibitem [{\citenamefont {Adam}\ \emph {et~al.}(2021)\citenamefont {Adam} \emph
  {et~al.}}]{STAR:2020tga}%
  \BibitemOpen
  \bibfield  {author} {\bibinfo {author} {\bibfnamefont {J.}~\bibnamefont
  {Adam}} \emph {et~al.} (\bibinfo {collaboration} {STAR}),\ }\href {\doibase
  10.1103/PhysRevLett.126.092301} {\bibfield  {journal} {\bibinfo  {journal}
  {Phys. Rev. Lett.}\ }\textbf {\bibinfo {volume} {126}},\ \bibinfo {pages}
  {092301} (\bibinfo {year} {2021})},\ \Eprint
  {http://arxiv.org/abs/2001.02852} {arXiv:2001.02852 [nucl-ex]} \BibitemShut
  {NoStop}%
\bibitem [{\citenamefont {Shen}\ and\ \citenamefont
  {Alzhrani}(2020)}]{Shen:2020jwv}%
  \BibitemOpen
  \bibfield  {author} {\bibinfo {author} {\bibfnamefont {C.}~\bibnamefont
  {Shen}}\ and\ \bibinfo {author} {\bibfnamefont {S.}~\bibnamefont
  {Alzhrani}},\ }\href {\doibase 10.1103/PhysRevC.102.014909} {\bibfield
  {journal} {\bibinfo  {journal} {Phys. Rev. C}\ }\textbf {\bibinfo {volume}
  {102}},\ \bibinfo {pages} {014909} (\bibinfo {year} {2020})},\ \Eprint
  {http://arxiv.org/abs/2003.05852} {arXiv:2003.05852 [nucl-th]} \BibitemShut
  {NoStop}%
\bibitem [{\citenamefont {Vovchenko}(2022)}]{Vovchenko:2021yen}%
  \BibitemOpen
  \bibfield  {author} {\bibinfo {author} {\bibfnamefont {V.}~\bibnamefont
  {Vovchenko}},\ }\href {\doibase 10.1103/PhysRevC.105.014903} {\bibfield
  {journal} {\bibinfo  {journal} {Phys. Rev. C}\ }\textbf {\bibinfo {volume}
  {105}},\ \bibinfo {pages} {014903} (\bibinfo {year} {2022})},\ \Eprint
  {http://arxiv.org/abs/2106.13775} {arXiv:2106.13775 [hep-ph]} \BibitemShut
  {NoStop}%
\bibitem [{\citenamefont {Adamczewski-Musch}\ \emph {et~al.}(2020)\citenamefont
  {Adamczewski-Musch} \emph {et~al.}}]{HADES:2020wpc}%
  \BibitemOpen
  \bibfield  {author} {\bibinfo {author} {\bibfnamefont {J.}~\bibnamefont
  {Adamczewski-Musch}} \emph {et~al.} (\bibinfo {collaboration} {HADES}),\
  }\href {\doibase 10.1103/PhysRevC.102.024914} {\bibfield  {journal} {\bibinfo
   {journal} {Phys. Rev. C}\ }\textbf {\bibinfo {volume} {102}},\ \bibinfo
  {pages} {024914} (\bibinfo {year} {2020})},\ \Eprint
  {http://arxiv.org/abs/2002.08701} {arXiv:2002.08701 [nucl-ex]} \BibitemShut
  {NoStop}%
\bibitem [{\citenamefont {Vovchenko}\ and\ \citenamefont
  {Koch}(2022)}]{Vovchenko:2022szk}%
  \BibitemOpen
  \bibfield  {author} {\bibinfo {author} {\bibfnamefont {V.}~\bibnamefont
  {Vovchenko}}\ and\ \bibinfo {author} {\bibfnamefont {V.}~\bibnamefont
  {Koch}},\ }\href {\doibase 10.1016/j.physletb.2022.137368} {\bibfield
  {journal} {\bibinfo  {journal} {Phys. Lett. B}\ } (\bibinfo {year} {2022}),\
  10.1016/j.physletb.2022.137368},\ \Eprint {http://arxiv.org/abs/2204.00137}
  {arXiv:2204.00137 [hep-ph]} \BibitemShut {NoStop}%
\bibitem [{\citenamefont {Harabasz}\ \emph {et~al.}(2020)\citenamefont
  {Harabasz}, \citenamefont {Florkowski}, \citenamefont {Galatyuk},
  \citenamefont {Ma~Lgorzata~Gumberidze}, \citenamefont {Ryblewski},
  \citenamefont {Salabura},\ and\ \citenamefont {Stroth}}]{Harabasz:2020sei}%
  \BibitemOpen
  \bibfield  {author} {\bibinfo {author} {\bibfnamefont {S.}~\bibnamefont
  {Harabasz}}, \bibinfo {author} {\bibfnamefont {W.}~\bibnamefont
  {Florkowski}}, \bibinfo {author} {\bibfnamefont {T.}~\bibnamefont
  {Galatyuk}}, \bibinfo {author} {\bibfnamefont {t.}~\bibnamefont
  {Ma~Lgorzata~Gumberidze}}, \bibinfo {author} {\bibfnamefont {R.}~\bibnamefont
  {Ryblewski}}, \bibinfo {author} {\bibfnamefont {P.}~\bibnamefont {Salabura}},
  \ and\ \bibinfo {author} {\bibfnamefont {J.}~\bibnamefont {Stroth}},\ }\href
  {\doibase 10.1103/PhysRevC.102.054903} {\bibfield  {journal} {\bibinfo
  {journal} {Phys. Rev. C}\ }\textbf {\bibinfo {volume} {102}},\ \bibinfo
  {pages} {054903} (\bibinfo {year} {2020})},\ \Eprint
  {http://arxiv.org/abs/2003.12992} {arXiv:2003.12992 [nucl-th]} \BibitemShut
  {NoStop}%
\bibitem [{\citenamefont {Motornenko}\ \emph
  {et~al.}(2021{\natexlab{a}})\citenamefont {Motornenko}, \citenamefont
  {Steinheimer}, \citenamefont {Vovchenko}, \citenamefont {Stock},\ and\
  \citenamefont {Stoecker}}]{Motornenko:2021nds}%
  \BibitemOpen
  \bibfield  {author} {\bibinfo {author} {\bibfnamefont {A.}~\bibnamefont
  {Motornenko}}, \bibinfo {author} {\bibfnamefont {J.}~\bibnamefont
  {Steinheimer}}, \bibinfo {author} {\bibfnamefont {V.}~\bibnamefont
  {Vovchenko}}, \bibinfo {author} {\bibfnamefont {R.}~\bibnamefont {Stock}}, \
  and\ \bibinfo {author} {\bibfnamefont {H.}~\bibnamefont {Stoecker}},\ }\href
  {\doibase 10.1016/j.physletb.2021.136703} {\bibfield  {journal} {\bibinfo
  {journal} {Phys. Lett. B}\ }\textbf {\bibinfo {volume} {822}},\ \bibinfo
  {pages} {136703} (\bibinfo {year} {2021}{\natexlab{a}})},\ \Eprint
  {http://arxiv.org/abs/2104.06036} {arXiv:2104.06036 [hep-ph]} \BibitemShut
  {NoStop}%
\bibitem [{\citenamefont {Abdallah}\ \emph {et~al.}(2022)\citenamefont
  {Abdallah} \emph {et~al.}}]{STAR:2021fge}%
  \BibitemOpen
  \bibfield  {author} {\bibinfo {author} {\bibfnamefont {M.~S.}\ \bibnamefont
  {Abdallah}} \emph {et~al.} (\bibinfo {collaboration} {STAR}),\ }\href
  {\doibase 10.1103/PhysRevLett.128.202303} {\bibfield  {journal} {\bibinfo
  {journal} {Phys. Rev. Lett.}\ }\textbf {\bibinfo {volume} {128}},\ \bibinfo
  {pages} {202303} (\bibinfo {year} {2022})},\ \Eprint
  {http://arxiv.org/abs/2112.00240} {arXiv:2112.00240 [nucl-ex]} \BibitemShut
  {NoStop}%
\bibitem [{\citenamefont {Ablyazimov}\ \emph {et~al.}(2017)\citenamefont
  {Ablyazimov} \emph {et~al.}}]{CBM:2016kpk}%
  \BibitemOpen
  \bibfield  {author} {\bibinfo {author} {\bibfnamefont {T.}~\bibnamefont
  {Ablyazimov}} \emph {et~al.} (\bibinfo {collaboration} {CBM}),\ }\href
  {\doibase 10.1140/epja/i2017-12248-y} {\bibfield  {journal} {\bibinfo
  {journal} {Eur. Phys. J. A}\ }\textbf {\bibinfo {volume} {53}},\ \bibinfo
  {pages} {60} (\bibinfo {year} {2017})},\ \Eprint
  {http://arxiv.org/abs/1607.01487} {arXiv:1607.01487 [nucl-ex]} \BibitemShut
  {NoStop}%
\bibitem [{\citenamefont {Bass}\ \emph {et~al.}(1998)\citenamefont {Bass} \emph
  {et~al.}}]{Bass:1998ca}%
  \BibitemOpen
  \bibfield  {author} {\bibinfo {author} {\bibfnamefont {S.~A.}\ \bibnamefont
  {Bass}} \emph {et~al.},\ }\href {\doibase 10.1016/S0146-6410(98)00058-1}
  {\bibfield  {journal} {\bibinfo  {journal} {Prog. Part. Nucl. Phys.}\
  }\textbf {\bibinfo {volume} {41}},\ \bibinfo {pages} {255} (\bibinfo {year}
  {1998})},\ \Eprint {http://arxiv.org/abs/nucl-th/9803035}
  {arXiv:nucl-th/9803035} \BibitemShut {NoStop}%
\bibitem [{\citenamefont {Bleicher}\ \emph {et~al.}(1999)\citenamefont
  {Bleicher} \emph {et~al.}}]{Bleicher:1999xi}%
  \BibitemOpen
  \bibfield  {author} {\bibinfo {author} {\bibfnamefont {M.}~\bibnamefont
  {Bleicher}} \emph {et~al.},\ }\href {\doibase 10.1088/0954-3899/25/9/308}
  {\bibfield  {journal} {\bibinfo  {journal} {J. Phys. G}\ }\textbf {\bibinfo
  {volume} {25}},\ \bibinfo {pages} {1859} (\bibinfo {year} {1999})},\ \Eprint
  {http://arxiv.org/abs/hep-ph/9909407} {arXiv:hep-ph/9909407} \BibitemShut
  {NoStop}%
\bibitem [{\citenamefont {Weil}\ \emph {et~al.}(2016)\citenamefont {Weil} \emph
  {et~al.}}]{Weil:2016zrk}%
  \BibitemOpen
  \bibfield  {author} {\bibinfo {author} {\bibfnamefont {J.}~\bibnamefont
  {Weil}} \emph {et~al.},\ }\href {\doibase 10.1103/PhysRevC.94.054905}
  {\bibfield  {journal} {\bibinfo  {journal} {Phys. Rev. C}\ }\textbf {\bibinfo
  {volume} {94}},\ \bibinfo {pages} {054905} (\bibinfo {year} {2016})},\
  \Eprint {http://arxiv.org/abs/1606.06642} {arXiv:1606.06642 [nucl-th]}
  \BibitemShut {NoStop}%
\bibitem [{\citenamefont {Sorensen}\ and\ \citenamefont
  {Koch}(2021)}]{Sorensen:2020ygf}%
  \BibitemOpen
  \bibfield  {author} {\bibinfo {author} {\bibfnamefont {A.}~\bibnamefont
  {Sorensen}}\ and\ \bibinfo {author} {\bibfnamefont {V.}~\bibnamefont
  {Koch}},\ }\href {\doibase 10.1103/PhysRevC.104.034904} {\bibfield  {journal}
  {\bibinfo  {journal} {Phys. Rev. C}\ }\textbf {\bibinfo {volume} {104}},\
  \bibinfo {pages} {034904} (\bibinfo {year} {2021})},\ \Eprint
  {http://arxiv.org/abs/2011.06635} {arXiv:2011.06635 [nucl-th]} \BibitemShut
  {NoStop}%
\bibitem [{\citenamefont {Savchuk}\ \emph {et~al.}(2022)\citenamefont
  {Savchuk}, \citenamefont {Vovchenko}, \citenamefont {Koch}, \citenamefont
  {Steinheimer},\ and\ \citenamefont {Stoecker}}]{Savchuk:2021aog}%
  \BibitemOpen
  \bibfield  {author} {\bibinfo {author} {\bibfnamefont {O.}~\bibnamefont
  {Savchuk}}, \bibinfo {author} {\bibfnamefont {V.}~\bibnamefont {Vovchenko}},
  \bibinfo {author} {\bibfnamefont {V.}~\bibnamefont {Koch}}, \bibinfo {author}
  {\bibfnamefont {J.}~\bibnamefont {Steinheimer}}, \ and\ \bibinfo {author}
  {\bibfnamefont {H.}~\bibnamefont {Stoecker}},\ }\href {\doibase
  10.1016/j.physletb.2022.136983} {\bibfield  {journal} {\bibinfo  {journal}
  {Phys. Lett. B}\ }\textbf {\bibinfo {volume} {827}},\ \bibinfo {pages}
  {136983} (\bibinfo {year} {2022})},\ \Eprint
  {http://arxiv.org/abs/2106.08239} {arXiv:2106.08239 [hep-ph]} \BibitemShut
  {NoStop}%
\bibitem [{\citenamefont {Garcia-Montero}\ \emph {et~al.}(2022)\citenamefont
  {Garcia-Montero}, \citenamefont {Staudenmaier}, \citenamefont {Sch\"afer},
  \citenamefont {Torres-Rincon},\ and\ \citenamefont
  {Elfner}}]{Garcia-Montero:2021haa}%
  \BibitemOpen
  \bibfield  {author} {\bibinfo {author} {\bibfnamefont {O.}~\bibnamefont
  {Garcia-Montero}}, \bibinfo {author} {\bibfnamefont {J.}~\bibnamefont
  {Staudenmaier}}, \bibinfo {author} {\bibfnamefont {A.}~\bibnamefont
  {Sch\"afer}}, \bibinfo {author} {\bibfnamefont {J.~M.}\ \bibnamefont
  {Torres-Rincon}}, \ and\ \bibinfo {author} {\bibfnamefont {H.}~\bibnamefont
  {Elfner}},\ }\href {\doibase 10.1103/PhysRevC.105.064906} {\bibfield
  {journal} {\bibinfo  {journal} {Phys. Rev. C}\ }\textbf {\bibinfo {volume}
  {105}} (\bibinfo {year} {2022}),\ 10.1103/PhysRevC.105.064906},\ \Eprint
  {http://arxiv.org/abs/2107.08812} {arXiv:2107.08812 [hep-ph]} \BibitemShut
  {NoStop}%
\bibitem [{FSg()}]{FSgithub}%
  \BibitemOpen
  \href@noop {} {}\bibinfo {note}
  {\href{https://github.com/vlvovch/fist-sampler}{https://github.com/vlvovch/fist-sampler}
  [Online; accessed 25-June-2021]}\BibitemShut {NoStop}%
\bibitem [{\citenamefont {Bernhard}(2020)}]{urqmd-afterburner-toolkit}%
  \BibitemOpen
  \bibfield  {author} {\bibinfo {author} {\bibfnamefont {J.}~\bibnamefont
  {Bernhard}},\ }\href@noop {} {\enquote {\bibinfo {title} {{UrQMD} tailored
  for use as a hadronic afterburner},}\ } (\bibinfo {year} {2020}),\ \bibinfo
  {note}
  {\href{https://github.com/jbernhard/urqmd-afterburner}{https://github.com/jbernhard/urqmd-afterburner}
  [Online; accessed 09-May-2021]}\BibitemShut {NoStop}%
\bibitem [{\citenamefont {Bass}\ \emph {et~al.}(2000)\citenamefont {Bass},
  \citenamefont {Danielewicz},\ and\ \citenamefont {Pratt}}]{Bass:2000az}%
  \BibitemOpen
  \bibfield  {author} {\bibinfo {author} {\bibfnamefont {S.~A.}\ \bibnamefont
  {Bass}}, \bibinfo {author} {\bibfnamefont {P.}~\bibnamefont {Danielewicz}}, \
  and\ \bibinfo {author} {\bibfnamefont {S.}~\bibnamefont {Pratt}},\ }\href
  {\doibase 10.1103/PhysRevLett.85.2689} {\bibfield  {journal} {\bibinfo
  {journal} {Phys. Rev. Lett.}\ }\textbf {\bibinfo {volume} {85}},\ \bibinfo
  {pages} {2689} (\bibinfo {year} {2000})},\ \Eprint
  {http://arxiv.org/abs/nucl-th/0005044} {arXiv:nucl-th/0005044} \BibitemShut
  {NoStop}%
\bibitem [{\citenamefont {Adam}\ \emph {et~al.}(2019)\citenamefont {Adam} \emph
  {et~al.}}]{STAR:2019ans}%
  \BibitemOpen
  \bibfield  {author} {\bibinfo {author} {\bibfnamefont {J.}~\bibnamefont
  {Adam}} \emph {et~al.} (\bibinfo {collaboration} {STAR}),\ }\href {\doibase
  10.1103/PhysRevC.100.014902} {\bibfield  {journal} {\bibinfo  {journal}
  {Phys. Rev. C}\ }\textbf {\bibinfo {volume} {100}},\ \bibinfo {pages}
  {014902} (\bibinfo {year} {2019})},\ \bibinfo {note} {[Erratum: Phys.Rev.C
  105, 029901 (2022)]},\ \Eprint {http://arxiv.org/abs/1903.05370}
  {arXiv:1903.05370 [nucl-ex]} \BibitemShut {NoStop}%
\bibitem [{\citenamefont {Everett}\ \emph
  {et~al.}(2021{\natexlab{a}})\citenamefont {Everett} \emph
  {et~al.}}]{JETSCAPE:2020shq}%
  \BibitemOpen
  \bibfield  {author} {\bibinfo {author} {\bibfnamefont {D.}~\bibnamefont
  {Everett}} \emph {et~al.} (\bibinfo {collaboration} {JETSCAPE}),\ }\href
  {\doibase 10.1103/PhysRevLett.126.242301} {\bibfield  {journal} {\bibinfo
  {journal} {Phys. Rev. Lett.}\ }\textbf {\bibinfo {volume} {126}},\ \bibinfo
  {pages} {242301} (\bibinfo {year} {2021}{\natexlab{a}})},\ \Eprint
  {http://arxiv.org/abs/2010.03928} {arXiv:2010.03928 [hep-ph]} \BibitemShut
  {NoStop}%
\bibitem [{\citenamefont {Everett}\ \emph
  {et~al.}(2021{\natexlab{b}})\citenamefont {Everett} \emph
  {et~al.}}]{JETSCAPE:2020mzn}%
  \BibitemOpen
  \bibfield  {author} {\bibinfo {author} {\bibfnamefont {D.}~\bibnamefont
  {Everett}} \emph {et~al.} (\bibinfo {collaboration} {JETSCAPE}),\ }\href
  {\doibase 10.1103/PhysRevC.103.054904} {\bibfield  {journal} {\bibinfo
  {journal} {Phys. Rev. C}\ }\textbf {\bibinfo {volume} {103}},\ \bibinfo
  {pages} {054904} (\bibinfo {year} {2021}{\natexlab{b}})},\ \Eprint
  {http://arxiv.org/abs/2011.01430} {arXiv:2011.01430 [hep-ph]} \BibitemShut
  {NoStop}%
\bibitem [{\citenamefont {Motornenko}\ \emph
  {et~al.}(2021{\natexlab{b}})\citenamefont {Motornenko}, \citenamefont {Pal},
  \citenamefont {Bhattacharyya}, \citenamefont {Steinheimer},\ and\
  \citenamefont {Stoecker}}]{Motornenko:2020yme}%
  \BibitemOpen
  \bibfield  {author} {\bibinfo {author} {\bibfnamefont {A.}~\bibnamefont
  {Motornenko}}, \bibinfo {author} {\bibfnamefont {S.}~\bibnamefont {Pal}},
  \bibinfo {author} {\bibfnamefont {A.}~\bibnamefont {Bhattacharyya}}, \bibinfo
  {author} {\bibfnamefont {J.}~\bibnamefont {Steinheimer}}, \ and\ \bibinfo
  {author} {\bibfnamefont {H.}~\bibnamefont {Stoecker}},\ }\href {\doibase
  10.1103/PhysRevC.103.054908} {\bibfield  {journal} {\bibinfo  {journal}
  {Phys. Rev. C}\ }\textbf {\bibinfo {volume} {103}},\ \bibinfo {pages}
  {054908} (\bibinfo {year} {2021}{\natexlab{b}})},\ \Eprint
  {http://arxiv.org/abs/2009.10848} {arXiv:2009.10848 [hep-ph]} \BibitemShut
  {NoStop}%
\bibitem [{\citenamefont {Sombun}\ \emph {et~al.}(2019)\citenamefont {Sombun},
  \citenamefont {Tomuang}, \citenamefont {Limphirat}, \citenamefont {Hillmann},
  \citenamefont {Herold}, \citenamefont {Steinheimer}, \citenamefont {Yan},\
  and\ \citenamefont {Bleicher}}]{Sombun:2018yqh}%
  \BibitemOpen
  \bibfield  {author} {\bibinfo {author} {\bibfnamefont {S.}~\bibnamefont
  {Sombun}}, \bibinfo {author} {\bibfnamefont {K.}~\bibnamefont {Tomuang}},
  \bibinfo {author} {\bibfnamefont {A.}~\bibnamefont {Limphirat}}, \bibinfo
  {author} {\bibfnamefont {P.}~\bibnamefont {Hillmann}}, \bibinfo {author}
  {\bibfnamefont {C.}~\bibnamefont {Herold}}, \bibinfo {author} {\bibfnamefont
  {J.}~\bibnamefont {Steinheimer}}, \bibinfo {author} {\bibfnamefont
  {Y.}~\bibnamefont {Yan}}, \ and\ \bibinfo {author} {\bibfnamefont
  {M.}~\bibnamefont {Bleicher}},\ }\href {\doibase 10.1103/PhysRevC.99.014901}
  {\bibfield  {journal} {\bibinfo  {journal} {Phys. Rev. C}\ }\textbf {\bibinfo
  {volume} {99}},\ \bibinfo {pages} {014901} (\bibinfo {year} {2019})},\
  \Eprint {http://arxiv.org/abs/1805.11509} {arXiv:1805.11509 [nucl-th]}
  \BibitemShut {NoStop}%
\bibitem [{\citenamefont {Hillmann}\ \emph {et~al.}(2022)\citenamefont
  {Hillmann}, \citenamefont {K\"afer}, \citenamefont {Steinheimer},
  \citenamefont {Vovchenko},\ and\ \citenamefont
  {Bleicher}}]{Hillmann:2021zgj}%
  \BibitemOpen
  \bibfield  {author} {\bibinfo {author} {\bibfnamefont {P.}~\bibnamefont
  {Hillmann}}, \bibinfo {author} {\bibfnamefont {K.}~\bibnamefont {K\"afer}},
  \bibinfo {author} {\bibfnamefont {J.}~\bibnamefont {Steinheimer}}, \bibinfo
  {author} {\bibfnamefont {V.}~\bibnamefont {Vovchenko}}, \ and\ \bibinfo
  {author} {\bibfnamefont {M.}~\bibnamefont {Bleicher}},\ }\href {\doibase
  10.1088/1361-6471/ac5dfc} {\bibfield  {journal} {\bibinfo  {journal} {J.
  Phys. G}\ }\textbf {\bibinfo {volume} {49}},\ \bibinfo {pages} {055107}
  (\bibinfo {year} {2022})},\ \Eprint {http://arxiv.org/abs/2109.05972}
  {arXiv:2109.05972 [hep-ph]} \BibitemShut {NoStop}%
\bibitem [{\citenamefont {Carnahan}\ and\ \citenamefont
  {Starling}(1969)}]{carnahan1969equation}%
  \BibitemOpen
  \bibfield  {author} {\bibinfo {author} {\bibfnamefont {N.~F.}\ \bibnamefont
  {Carnahan}}\ and\ \bibinfo {author} {\bibfnamefont {K.~E.}\ \bibnamefont
  {Starling}},\ }\href {\doibase 10.1063/1.1672048} {\bibfield  {journal}
  {\bibinfo  {journal} {The Journal of Chemical Physics}\ }\textbf {\bibinfo
  {volume} {51}},\ \bibinfo {pages} {635} (\bibinfo {year} {1969})}\BibitemShut
  {NoStop}%
\bibitem [{\citenamefont {Wu}\ and\ \citenamefont {Sadus}(2005)}]{wu2005hard}%
  \BibitemOpen
  \bibfield  {author} {\bibinfo {author} {\bibfnamefont {G.-W.}\ \bibnamefont
  {Wu}}\ and\ \bibinfo {author} {\bibfnamefont {R.~J.}\ \bibnamefont {Sadus}},\
  }\href {\doibase https://doi.org/10.1002/aic.10233} {\bibfield  {journal}
  {\bibinfo  {journal} {AIChE Journal}\ }\textbf {\bibinfo {volume} {51}},\
  \bibinfo {pages} {309} (\bibinfo {year} {2005})}\BibitemShut {NoStop}%
\bibitem [{\citenamefont {Wertheim}(1963)}]{PhysRevLett.10.321}%
  \BibitemOpen
  \bibfield  {author} {\bibinfo {author} {\bibfnamefont {M.~S.}\ \bibnamefont
  {Wertheim}},\ }\href {\doibase 10.1103/PhysRevLett.10.321} {\bibfield
  {journal} {\bibinfo  {journal} {Phys. Rev. Lett.}\ }\textbf {\bibinfo
  {volume} {10}},\ \bibinfo {pages} {321} (\bibinfo {year} {1963})}\BibitemShut
  {NoStop}%
\bibitem [{\citenamefont {Thiele}(1963)}]{doi:10.1063/1.1734272}%
  \BibitemOpen
  \bibfield  {author} {\bibinfo {author} {\bibfnamefont {E.}~\bibnamefont
  {Thiele}},\ }\href {\doibase 10.1063/1.1734272} {\bibfield  {journal}
  {\bibinfo  {journal} {The Journal of Chemical Physics}\ }\textbf {\bibinfo
  {volume} {39}},\ \bibinfo {pages} {474} (\bibinfo {year} {1963})}\BibitemShut
  {NoStop}%
\bibitem [{\citenamefont {Yuste}\ and\ \citenamefont
  {Santos}(1991)}]{PhysRevA.43.5418}%
  \BibitemOpen
  \bibfield  {author} {\bibinfo {author} {\bibfnamefont {S.~B.}\ \bibnamefont
  {Yuste}}\ and\ \bibinfo {author} {\bibfnamefont {A.}~\bibnamefont {Santos}},\
  }\href {\doibase 10.1103/PhysRevA.43.5418} {\bibfield  {journal} {\bibinfo
  {journal} {Phys. Rev. A}\ }\textbf {\bibinfo {volume} {43}},\ \bibinfo
  {pages} {5418} (\bibinfo {year} {1991})}\BibitemShut {NoStop}%
\bibitem [{\citenamefont {Kuznietsov}\ \emph {et~al.}(2022)\citenamefont
  {Kuznietsov}, \citenamefont {Savchuk}, \citenamefont {Gorenstein},
  \citenamefont {Koch},\ and\ \citenamefont {Vovchenko}}]{Kuznietsov:2022pcn}%
  \BibitemOpen
  \bibfield  {author} {\bibinfo {author} {\bibfnamefont {V.~A.}\ \bibnamefont
  {Kuznietsov}}, \bibinfo {author} {\bibfnamefont {O.}~\bibnamefont {Savchuk}},
  \bibinfo {author} {\bibfnamefont {M.~I.}\ \bibnamefont {Gorenstein}},
  \bibinfo {author} {\bibfnamefont {V.}~\bibnamefont {Koch}}, \ and\ \bibinfo
  {author} {\bibfnamefont {V.}~\bibnamefont {Vovchenko}},\ }\href {\doibase
  10.1103/PhysRevC.105.044903} {\bibfield  {journal} {\bibinfo  {journal}
  {Phys. Rev. C}\ }\textbf {\bibinfo {volume} {105}},\ \bibinfo {pages}
  {044903} (\bibinfo {year} {2022})},\ \Eprint
  {http://arxiv.org/abs/2201.08486} {arXiv:2201.08486 [hep-ph]} \BibitemShut
  {NoStop}%
\end{thebibliography}%

%TC:endignore 

\end{document}